\documentclass{article}
\usepackage[T1]{fontenc}
\usepackage{microtype}
\usepackage{graphicx}
\usepackage{booktabs}
\usepackage{amsmath,amssymb}
\usepackage{xcolor}
\usepackage{xurl}
\usepackage{hyperref}
\usepackage[accepted]{mlsys2025}
\makeatletter\renewcommand{\Notice@String}{Preprint.}\makeatother
\AtBeginDocument{%
  \setlength{\dbltextfloatsep}{11pt plus 2pt minus 2pt}%
  \setlength{\textfloatsep}{11pt plus 2pt minus 2pt}%
}
\hypersetup{pdfsubject={},pdfauthor={Yichao Fu, Yuxuan Zhang, Minshen Zhang, Ruitian Wang, Junda Chen, Hao Zhang}}
\newcommand{\sys}{FastAFD}
\newcommand{\qwen}{Qwen3}
\newcommand{\qwenfull}{Qwen3-235B-A22B}
\newcommand{\hybrid}{Qwen3.5}
\newcommand{\hybridfull}{Qwen3.5-397B-A17B}
\newcommand{\mm}{MiniMax-M2.5}

\newcommand{\revisionNov}[1]{#1}
\makeatletter\newcommand{\revisionNovFig}{\renewcommand{\fnum@figure}{\revisionNov{\figurename~\thefigure}}}\makeatother

\newcommand{\allgpu}{\mathrm{all}}
\newcommand{\attgpu}{\mathrm{att}}
\newcommand{\specmode}{\mathrm{spec}}
\newcommand{\inputoptional}[1]{\IfFileExists{#1}{\input{#1}}{\textbf{Missing generated table: #1}}}

\mlsystitlerunning{FastAFD: When Does Attention--FFN Disaggregation Pay?}
\begin{document}
\raggedbottom
\twocolumn[
\mlsystitle{FastAFD: When Does Attention-FFN Disaggregation Pay on Rack-Scale NVLink?}
\hypersetup{pdftitle={FastAFD: When Does Attention-FFN Disaggregation Pay on Rack-Scale NVLink?}}%
\mlsyssetsymbol{equal}{*}
\begin{mlsysauthorlist}
\mlsysauthor{Yichao Fu}{equal,ucsd}
\mlsysauthor{Yuxuan Zhang}{equal,ucsd}
\mlsysauthor{Minshen Zhang}{equal,ucsd}
\mlsysauthor{Ruitian Wang}{ucsd}
\mlsysauthor{Junda Chen}{ucsd}
\mlsysauthor{Hao Zhang}{ucsd}
\end{mlsysauthorlist}
\mlsysaffiliation{ucsd}{University of California, San Diego}
\mlsyscorrespondingauthor{Hao Zhang}{haozhang@ucsd.edu}
\mlsyskeywords{Mixture-of-Experts, inference, disaggregation, speculative decoding}
\vskip 0.3in
\begin{abstract}
Attention--FFN disaggregation (AFD) separates attention from the experts of a mixture-of-experts (MoE) model and has been reported to raise decode throughput over colocated serving.
Whether this advantage holds on rack-scale NVLink systems such as GB200 NVL72 remains unclear: prior results were obtained on other hardware, against narrower colocated baselines, or with analytical models, whereas production engines serve large MoE models on these systems with wide expert parallelism (wide EP).
We build and carefully tune \sys{}, an AFD inference engine for GB200 NVL72 that overlaps the MoE of one microbatch with the attention of the other, and compare it against the best vLLM wide EP setting (with the EP width tuned per workload).
Our experiments show that AFD has a throughput advantage of up to \revisionNov{$1.17\times$ over wide EP on \qwen{} and $1.16\times$ on \hybrid{}, and matches it on \mm{}}.
We derive an analytical model that predicts AFD throughput within $3\%$ using only profiles of colocated vLLM, which tells whether switching to AFD would improve throughput before AFD is deployed.
Beyond prediction, we turn the model into a concise insight: AFD's speedup factors into a batch gain (larger batches, since attention GPUs hold no expert weights), an overlap gain (shorter steps, since the MoE runs on the expert GPUs behind attention), and a disaggregation cost (the GPUs given to experts). AFD has the advantage when the two gains outweigh the cost. This view explains which models benefit from AFD, and shows that speculative decoding helps light-expert models most, where the enlarged overlap gain outweighs the larger disaggregation cost and adds a further $8\%$ to AFD's advantage over wide EP.
Looking ahead, serving the experts on accelerators that keep their weights in on-chip SRAM, such as the Groq LPX accelerators that NVIDIA pairs with Vera Rubin NVL72, would remove most of the disaggregation cost, raising the ideal speedup to $2.48\times$ with speculative decoding.
\sys{} is open source at \url{https://github.com/hao-ai-lab/FastAFD}.
\end{abstract}
]
\printAffiliationsAndNotice{\mlsysEqualContribution}
\section{Introduction}

Attention--FFN disaggregation (AFD) has drawn sharply divided conclusions as a design for large-scale mixture-of-experts (MoE)\revisionNov{~\cite{shazeer2017,gshard,switch}} LLM serving: some works report substantial decode-throughput gains over colocated deployment~\cite{megascale,step3}, while others argue that it cannot match a well-tuned wide expert-parallel (wide EP) baseline~\cite{afdledger}. 

Which design serves MoE models better on rack-scale NVLink systems remains unclear. On one hand, AFD separates attention from experts: attention GPUs hold the KV caches and they run larger decode batches, while expert GPUs gather tokens from attention GPUs into one batch, therefore the MoE GEMMs run closer to their roofline. MegaScale-Infer reports up to $1.90\times$ higher per-GPU decode throughput than colocated serving baselines~\cite{megascale}, Step-3 co-designs its model around AFD~\cite{step3}, NPU superpods deploy it at scale~\cite{xdeepserve}, NVIDIA builds AFD into its Vera Rubin platform by running MoE experts on Groq LPX accelerators~\cite{nvidiagroqlpx}, and vLLM ships an experimental AFD plugin~\cite{vllmafd}. On the other hand, wide EP~\cite{dsmoe,deepseekv3} keeps attention and experts on every GPU but shards the experts across a rack-scale NVLink domain such as GB200 NVL72~\cite{gb200}, which frees much of the same HBM for KV caches; vLLM and SGLang serve large MoE models this way~\cite{vllmwideep,vllmgb200,sglanggb200}, and analytical studies of modern hardware lean the same way: a communication-roofline analysis finds AFD promising only for specific hardware-model combinations~\cite{afdchallenges}, a design-space exploration on a 128-GPU B200 cluster finds aggregated serving keeps the throughput lead~\cite{howfar}, and a provisioning model that sweeps all EP widths finds AFD faster in only 7 of 36 settings~\cite{afdledger}. The measured evidence for AFD, however, comes from older hardware against narrow colocated baselines, and the evidence against it is analytical and unvalidated on a rack-scale NVLink system. As a result, there is no measured basis for choosing between the two designs on GB200 NVL72, a platform on which large MoE models are now served~\cite{sglanggb200,vllmgb200}.

\begin{figure}[t]
\centering
\includegraphics[width=\columnwidth]{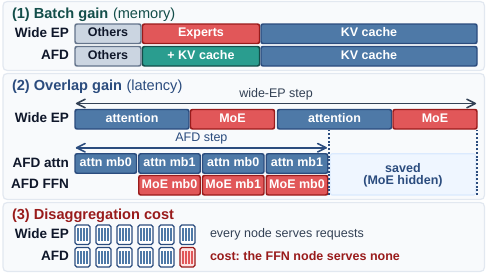}
\vspace{-1.6em}
\caption{\textbf{AFD's speedup over wide EP balances two gains against one cost (Eq.~\ref{eq:identity}).} \textbf{(1) Batch gain}: moving expert weights off attention GPUs frees HBM for larger KV caches and decode batches. \textbf{(2) Overlap gain}: two-microbatch ping-pong hides each microbatch's MoE behind the other's attention stage. \textbf{(3) Disaggregation cost}: attention nodes hold KV caches and serve requests; the FFN node does not, so one node in every attention:FFN group produces no tokens.}
\label{fig:intro-balance}
\end{figure}

To close this gap, we build \sys{}, an AFD inference engine for GB200 NVL72 whose kernels, microbatching, and coordination are each tuned for the rack (Section~\ref{sec:system}), and benchmark it against vLLM wide EP with the EP width tuned per workload, on three MoE models. We show empirically that AFD outperforms this baseline by up to \revisionNov{$1.17\times$ (\qwen{}-235B at 16K context)}. Based on this empirical study, we further derive an analytical model that predicts AFD throughput using only profiles of the colocated vLLM deployment; its predictions stay within $3\%$ of the measured throughput on average. Beyond prediction, we turn the analytical model into a concise key insight: AFD's speedup over wide EP is the product of two gains and one cost (Figure~\ref{fig:intro-balance}):
\begin{equation}
S_{\allgpu} \;=\; \text{batch gain}\times\text{overlap gain}\times\text{disaggregation cost}.
\label{eq:identity}
\end{equation}
(1) The \emph{batch gain} comes from attention GPUs that hold no expert weights and thus fit a larger KV cache; (2) the \emph{overlap gain} comes from taking the MoE off the attention critical path; and (3) the \emph{disaggregation cost} comes from the expert GPUs, which serve no requests. AFD has the advantage when the two gains outweigh the cost.

This insight also predicts when speculative decoding helps AFD. On \mm{}, whose light experts yield a small overlap gain, AFD only ties wide EP. Speculative decoding multiplies the MoE work on this model, which enlarges the overlap gain, while the disaggregation cost remains small and therefore lifts advantage by $8\%$.

Looking ahead, \emph{heterogeneous FFN placement} on dedicated SRAM accelerators such as Groq LPX paired with Vera Rubin NVL72~\cite{nvidiagroqlpx} (Section~\ref{sec:provisioning}) removes the disaggregation cost from the GPU budget, raising AFD's best speedup over wide EP from \revisionNov{$1.17\times$} with homogeneous AFD to an ideal $2.48\times$ with heterogeneous placement and speculative decoding.

In summary, this paper mainly makes three contributions:
\begin{itemize}
\item We build and carefully tune \sys{}, the first AFD inference engine for GB200 NVL72, and benchmark it against the best wide EP baselines, which it consistently matches or outperforms with up to a \revisionNov{$1.17\times$} throughput advantage (Sections~\ref{sec:baseline} and~\ref{sec:system}).
\item We derive an analytical model that predicts \sys{}'s throughput within $3\%$ mean error. It needs only a profile of vLLM on the target cluster to tell how much switching to AFD would speed up a deployment, without actually deploying AFD (Section~\ref{sec:model}).
\item We turn the model into a concise insight with two gains (batch and overlap) and one disaggregation cost. The insight guides AFD tuning decisions: which models benefit from AFD, when speculative decoding helps, and how much headroom heterogeneous FFN placement unlocks (Sections~\ref{sec:model} and~\ref{sec:provisioning}).
\end{itemize}

\section{Background and Motivation}
\label{sec:background}

\begin{figure*}[t]
\centering
\revisionNovFig\includegraphics[width=\textwidth]{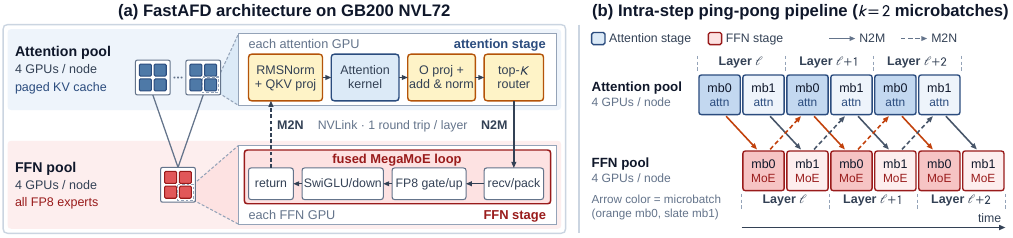}
\vspace{-1.15em}
\caption{\textbf{\sys{} architecture and intra-step ping-pong overlap on GB200 NVL72.} \textbf{(a)} Attention nodes (blue) hold paged KV caches and run the attention stage (norms, QKV/O projections, TensorRT-LLM FMHA attention, and top-$K$ routing), while one NVLink round trip per layer (\texttt{N2M}/\texttt{M2N}) dispatches routed tokens to a dedicated FFN node (coral) running a single persistent fused \texttt{MegaMoE} kernel. \textbf{(b)} Splitting each attention GPU's batch into two ping-pong microbatches (\texttt{mb0}, \texttt{mb1}) overlaps the attention stage on the attention pool with the FFN stage (\texttt{MegaMoE}) on the FFN pool across consecutive layers.}
\vspace{-0.75em}
\label{fig:arch}
\label{fig:pipeline}
\end{figure*}

\subsection{Colocated Wide EP and Attention--FFN Disaggregation}
MoE decoding couples stateful attention with stateless expert feed-forward networks (FFNs). During each decode step, attention reads every active request's growing key--value (KV) cache from fixed per-GPU HBM, whereas expert FFNs hold no per-request state and stream shared expert weights across requests. In colocated serving~\cite{vllm,sglang}, every GPU stores a full copy of the attention weights alongside a shard of the routed experts. Single-node expert parallelism (e.g., \texttt{EP4}) keeps one quarter of all experts on each GPU, leaving limited HBM for KV caches and capping the per-GPU decode batch. On a 72-GPU GB200 NVL72 rack~\cite{gb200}, \emph{colocated wide EP} (e.g., \texttt{EP8}--\texttt{EP64})~\cite{dsmoe,deepseekv3,vllmwideep,sglanggb200} shards experts across 8--64 GPUs while keeping attention data-parallel. Shrinking each GPU's expert shard frees HBM for larger decode batches, while each step still executes attention, NVLink all-to-all routing, and expert FFN serially on the same GPU.

Rather than colocating both modules, \revisionNov{AFD}~\cite{megascale,step3} separates MoE inference into an \emph{attention pool} and a dedicated \emph{FFN pool} provisioned at a chosen \emph{A:F node ratio} (Figure~\ref{fig:arch}a). Because attention GPUs store no expert weights, they dedicate all free HBM to KV caches and maximize their per-GPU decode batch. Within a single batch, however, layer $\ell{+}1$'s attention depends sequentially on the return of layer $\ell$'s remote experts (\texttt{N2M} dispatch $\rightarrow$ experts $\rightarrow$ \texttt{M2N} combine), which would serialize the two pools. To overlap them, AFD splits each attention GPU's batch into two ping-pong microbatches (\texttt{mb0} and \texttt{mb1} in Figure~\ref{fig:arch}b): in each layer, while the FFN pool runs \texttt{mb0}'s \emph{FFN stage}, the attention pool concurrently runs \texttt{mb1}'s \emph{attention stage} (norms, projections, core attention, and routing), and the two microbatches then swap roles.

\subsection{Motivation: Why Re-evaluate AFD on Rack-Scale NVLink?}

\textbf{NVL72 strengthens the baseline as much as AFD.}
With $1.8\,\text{TB/s}$ NVLink to each of 72 GPUs, remote expert dispatch costs no more than an intra-node hop, which enables rack-scale AFD and the heterogeneous racks built around it~\cite{nvidiagroqlpx}. The same fabric lets colocated serving shard experts across the rack (\texttt{EP8}--\texttt{EP64}) and reclaim most of the HBM that AFD frees, so the \texttt{TP} and \texttt{EP4} baselines of prior AFD studies~\cite{megatron,megascale} no longer represent the best colocated deployment. The comparison must also charge AFD for its FFN nodes, which hold no KV cache and serve no requests.

\textbf{The existing answers for NVL72 are models, and they disagree.}
The two studies that address rack-scale NVLink directly are analytical: one finds AFD likely to benefit only on superpod-class interconnects and models with coarse-grained experts~\cite{afdchallenges}, the other finds homogeneous AFD ahead in only 7 of 36 settings~\cite{afdledger}. Neither is checked against a rack-scale measurement, and MegaScale-Infer's $T_a \approx T_e$ rule picks AFD's ratio but has no wide-EP counterpart~\cite{megascale}.

\textbf{Speculative decoding is standard in production and changes the question.}
Speculative decoding~\cite{leviathan,chen,eagle} is supported by every major serving engine and accelerates decoding across production workloads~\cite{sdillusion}; frontier MoE models ship multi-token-prediction heads for it~\cite{deepseekv3}. Verifying $q$ draft tokens per step reuses each request's KV-cache read but multiplies routed expert work by $q$: under wide EP that work lands on every GPU's critical path, under AFD on an FFN pool that standard decoding leaves partly idle. No study has measured AFD under speculation.

We answer all three on hardware (Section~\ref{sec:results}), with an engine built to add no slack of its own (Section~\ref{sec:overview}) and a model that covers both designs (Section~\ref{sec:model}).
\subsection{\sys{} Overview}\label{sec:overview}
Such a runtime is harder to build on GB200 NVL72 than on earlier clusters. Because $7.9\,\text{TB/s}$ HBM3e finishes a layer's KV reads and expert-weight streaming in a few hundred microseconds, anything that touches the host inside a decode step is visible in the measurement. Prior AFD systems were built for multi-node clusters connected over InfiniBand or RoCE, where a layer takes on the order of a millisecond, and they keep the CPU in the per-layer communication loop: MegaScale-Infer's M2N library issues RDMA writes and polls completions from host threads~\cite{megascale}, and Step-3's StepMesh drives GPUDirect RDMA from CPU network threads so that communication consumes no GPU SMs~\cite{step3}. On NVL72 the same host round trip is a visible fraction of the step. The runtime must also be comparable: any difference in attention kernels or numeric precision between the two engines would be mistaken for an architectural difference.

Five design choices follow (Figure~\ref{fig:arch}a). (1) A MoE-boundary split with matched kernels makes the comparison isolate architecture rather than kernels: built on Mini-SGLang~\cite{minisglang}, \sys{} keeps norms, QKV/O projections, and top-$K$ routing on attention GPUs and uses colocated vLLM's exact attention kernels (TensorRT-LLM FMHA~\cite{trtllm}) and block-FP8 expert precision~\cite{deepseekv3}. The other four remove the four places where a decode step would otherwise wait: on the host within a layer, on a launch between kernels, on communication between pools, and on the host between steps. (2) Within a layer, GPU-initiated, one-sided communication replaces host posts and polls: attention GPUs publish FP8 tokens and routing metadata into pre-allocated symmetric NVLink buffers, at the cost of 24 SMs on each attention GPU for the fused send/receive kernel. (3) Between dispatch, GEMM, and combine, a persistent fused \texttt{MegaMoE} kernel on FFN GPUs replaces the separate launches of modular pipelines~\cite{deepep,deepgemm}, fusing the NVLink TMA pull (\texttt{N2M}), block-FP8 grouped GEMMs, activation, and push-back (\texttt{M2N}) into one launch per layer. (4) Across the two pools, two ping-pong microbatches ($k=2$) hide remote \texttt{MegaMoE} execution behind local attention: NVL72's bandwidth covers the round trip with two microbatches, whereas slower interconnects require $k \ge 3$~\cite{megascale} and pay for it in extra dense-projection passes and smaller GEMM tiles. (5) Between steps, whole-step CUDA graph capture with a one-step-ahead CPU coordinator replaces synchronous host bookkeeping: because (2) leaves no host synchronization inside the step, the whole step replays as one graph, and the host prepares Step $t{+}1$ (batch composition, padding, buffer slots, KV-cache frees) while Step $t$ runs. Each choice is ablated, alongside alternative stage boundaries~\cite{distca}, in Section~\ref{sec:system-ablation}. Because only attention GPUs hold KV caches and emit tokens, our evaluation reports both \emph{all-GPU speedup} (charging all allocated attention and FFN GPUs) and \emph{attention-GPU speedup} (counting attention GPUs only, which reflects per-GPU gain when the FFN pool is placed off-GPU).
\section{End-to-End Performance on NVL72}
\label{sec:baseline}

\begin{figure*}[t]
\centering
\revisionNovFig\includegraphics[width=\textwidth]{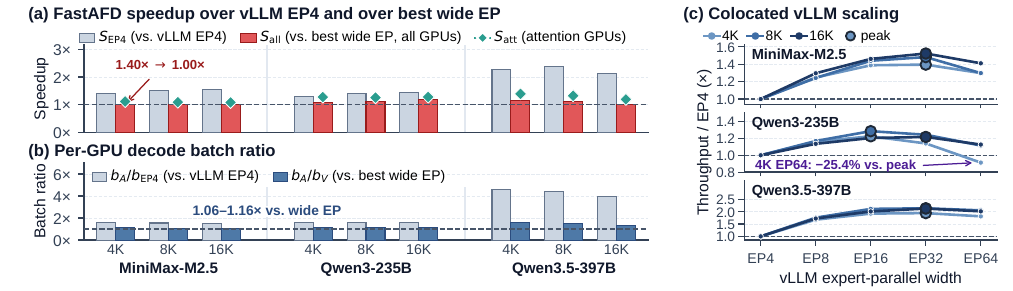}
\vspace{-1.15em}
\caption{\textbf{Context-tuned wide EP narrows homogeneous AFD gains from \revisionNov{$1.30\text{--}2.36\times$} to \revisionNov{$0.99\text{--}1.17\times$} across 4K--16K contexts.} \textbf{(a)} \sys{} speedups over EP4 ($S_{\mathrm{EP4}}$, slate) and over best vLLM wide EP counting all GPUs ($S_{\allgpu}$, coral) or attention GPUs only ($S_{\attgpu}$, teal; stems show the disaggregation cost). \textbf{(b)} Per-GPU batch ratios ($b_A/b_{\mathrm{EP4}}$ vs.\ $b_A/b_V$). \textbf{(c)} Colocated vLLM scaling across EP4--EP64 normalized to EP4 (outlined markers: per-curve optima).}
\vspace{-0.85em}
\label{fig:overview}
\label{fig:main}
\end{figure*}

\subsection{Experiment Setup}\label{sec:setup}
We evaluate inference decoding on a single GB200 NVL72 rack comprising 18 four-GPU nodes within one 72-GPU NVLink domain~\cite{gb200}. We use three production open-weight FP8 MoE models~\cite{minimaxm25,qwen3,qwen35} chosen to cover the three architectural axes that govern AFD's trade-off: light-expert full attention (\mm{}, 10B active of 229B across 256 experts), heavy-expert full attention (Qwen3-235B~\cite{qwen3tr}, 22B active of 235B across 128 experts), and ultra-sparse hybrid linear-plus-full attention (Qwen3.5-397B, 17B active of 397B across 512 experts; full specifications in Appendix~A).

\textbf{Workloads and protocol.}
Each request carries a fixed prompt of 4K, 8K, or 16K tokens (Ctx in Tables~\ref{tab:std} and~\ref{tab:spec}). We prefill once, then measure steady-state decode throughput in closed loop with every slot occupied, over a window short enough that per-request KV length stays close to the nominal context; prefill is excluded, since AFD is a decode-phase design and prefill is served by prefill--decode disaggregation. Both engines reserve 2\,GB of HBM for activations and communication buffers and give all remaining capacity to the BF16 paged KV cache, running the largest decode batch that fits it; per-GPU batch is therefore set by KV capacity, and because a colocated vLLM GPU holds $1/\texttt{EP}$ of the expert weights beside its KV cache while a \sys{} attention GPU holds none, the difference in free HBM is the batch gain $b_A/b_V$ of Section~\ref{sec:model-decomp}.

\textbf{Configurations.}
For each workload, colocated vLLM~\cite{vllm} scales expert parallelism across nodes from single-node \texttt{EP4} through \texttt{EP8}, \texttt{EP16}, \texttt{EP32}, and \texttt{EP64} (\texttt{TP1}, \texttt{DP}$=$\texttt{EP}), while \sys{} dedicates one four-GPU node to the FFN pool and scales the attention-to-FFN node ratio $r = N_A / N_F$ over ratios from 1:1 to 17:1 until per-GPU decode throughput peaks and the next wider step turns down, or until reaching the 18-node rack limit. Each engine is reported at its highest-throughput configuration per workload.

\textbf{Speculative decoding.}
We evaluate draft depth $d=3$ ($q=4$ verified tokens per step) on \mm{} and Qwen3-235B. Draft tokens are simulated with acceptance fixed at $\bar a\approx 1.8$ accepted tokens per step for both engines; since both engines verify the same $q$ tokens with the same acceptance, the comparison depends on how each serves the multiplied expert work, not on draft quality.

\textbf{Metrics.}
Our primary metric, all-GPU speedup ($S_{\allgpu}$), divides total decode throughput by every allocated GPU (\texttt{EP} GPUs for vLLM, $4(r{+}1)$ for \sys{}), charging \sys{} for its dedicated FFN nodes; we also report $S_{\mathrm{EP4}}$ against single-node \texttt{EP4}. We pair $S_{\allgpu}$ with attention-GPU speedup ($S_{\attgpu} = \frac{r+1}{r}S_{\allgpu}$), which counts only attention GPUs and measures the per-GPU gain when the FFN pool is placed on non-GPU accelerators. To rule out kernel discrepancies, both engines use identical TensorRT-LLM FMHA attention kernels~\cite{trtllm} and DeepSeek-V3 block-FP8 expert precision~\cite{deepseekv3}.

\begin{figure*}[t]
\centering
\includegraphics[width=\textwidth]{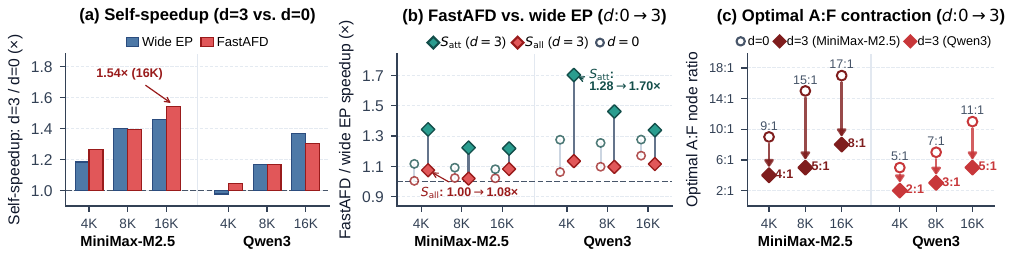}
\vspace{-1.15em}
\caption{\textbf{Speculative decoding ($d=3, q=4, \bar a\approx 1.8$) raises $S_{\attgpu}$ on every workload and keeps $S_{\allgpu}$ at $1.02\text{--}1.13\times$ while contracting optimal A:F ratios (4K--16K).} \textbf{(a)} Same-engine speculative self-speedup over $d=0$ (\sys{} up to $1.54\times$). \textbf{(b)} \sys{} speedup over tuned wide EP ($d=0$ circles vs.\ $d=3$ diamonds): hiding $4\times$ MoE work on the FFN pool raises $S_{\attgpu}$ on all six workloads (up to $1.70\times$) and lifts \mm{} $S_{\allgpu}$ to $1.08\times$ at 4K and 16K. \textbf{(c)} Optimal A:F ratios contract ($d\text{: }0\rightarrow 3$) as the FFN pool takes $rnq$ routed tokens.}
\vspace{-0.85em}
\label{fig:spec}
\end{figure*}

\subsection{Results}
\label{sec:results}
\begin{table}[t]
\caption{\textbf{Standard decoding ($d=0$): context-tuned wide EP erases most of \sys{}'s gain over EP4.} Each engine runs at its measured best configuration: the vLLM EP width (Wide EP) and the \sys{} attention:FFN node ratio (A:F). $S_{\mathrm{EP4}}$ and $S_{\allgpu}$ compare per-GPU decode throughput over EP4 and the best wide EP, counting every allocated GPU; $S_{\attgpu}$ counts attention GPUs only. $\blacktriangle$/$\blacktriangledown$: largest/smallest in each column; $\dagger$: still rising at the 18-node rack limit (a lower bound).}
\label{tab:std}\centering\scriptsize
\setlength{\tabcolsep}{2.8pt}
\begin{tabular}{llcrrrr}\toprule
 & & \multicolumn{2}{c}{Best configuration} & \multicolumn{3}{c}{Speedup ($\times$)}\\
\cmidrule(lr){3-4}\cmidrule(lr){5-7}
Model & Ctx & Wide EP & A:F & $S_{\mathrm{EP4}}$ & $S_{\allgpu}$ & $S_{\attgpu}$\\\midrule
\mm{} & 4K & EP32 & 9:1 & 1.40 & \revisionNov{1.00} & 1.12\\
 & 8K & EP32 & 15:1 & 1.52 & 1.02 & 1.09\\
 & 16K & EP32 & 17:1$^{\dagger}$ & 1.56 & 1.02 & 1.08$^{\scriptscriptstyle\blacktriangledown}$\\\midrule
\qwen{} & 4K & EP16 & 5:1 & 1.30$^{\scriptscriptstyle\blacktriangledown}$ & 1.06 & 1.28\\
 & 8K & EP16 & 7:1 & 1.41 & 1.10 & 1.26\\
 & 16K & EP32 & 11:1 & 1.42 & \revisionNov{1.17$^{\scriptscriptstyle\blacktriangle}$} & 1.28\\\midrule
\hybrid{} & 4K & EP32 & 5:1 & \revisionNov{2.25} & \revisionNov{1.16} & \revisionNov{1.39$^{\scriptscriptstyle\blacktriangle}$}\\
 & 8K & EP32 & 5:1 & \revisionNov{2.36$^{\scriptscriptstyle\blacktriangle}$} & \revisionNov{1.10} & \revisionNov{1.33}\\
 & 16K & EP32 & 5:1 & \revisionNov{2.11} & \revisionNov{0.99$^{\scriptscriptstyle\blacktriangledown}$} & \revisionNov{1.19}\\
\bottomrule
\end{tabular}
\vspace{-0.85em}
\end{table}
\begin{table}[t]
\caption{\textbf{Speculative decoding ($d=3$) at least halves \sys{}'s best A:F ratio and raises $S_{\attgpu}$ on every workload.} Acceptance is held near $\bar a\approx 1.8$ tokens per step. A:F shows the best ratio at $d=0\rightarrow d=3$; Wide EP is the best vLLM width at $d=3$. Speedups as in Table~\ref{tab:std}.}
\label{tab:spec}\centering\scriptsize
\setlength{\tabcolsep}{4pt}
\begin{tabular}{llccrr}\toprule
 & & \multicolumn{2}{c}{Best configuration} & \multicolumn{2}{c}{Speedup ($\times$)}\\
\cmidrule(lr){3-4}\cmidrule(lr){5-6}
Model & Ctx & Wide EP & A:F & $S_{\allgpu}$ & $S_{\attgpu}$\\\midrule
\mm{} & 4K & EP16 & 9:1$\,\rightarrow\,$4:1 & 1.08 & 1.34\\
 & 8K & EP16 & 15:1$\,\rightarrow\,$5:1 & 1.02$^{\scriptscriptstyle\blacktriangledown}$ & 1.22\\
 & 16K & EP16 & 17:1$\,\rightarrow\,$8:1 & 1.08 & 1.22$^{\scriptscriptstyle\blacktriangledown}$\\\midrule
\qwen{} & 4K & EP8 & 5:1$\,\rightarrow\,$2:1 & 1.13$^{\scriptscriptstyle\blacktriangle}$ & 1.70$^{\scriptscriptstyle\blacktriangle}$\\
 & 8K & EP16 & 7:1$\,\rightarrow\,$3:1 & 1.10 & 1.46\\
 & 16K & EP8 & 11:1$\,\rightarrow\,$5:1 & 1.12 & 1.34\\
\bottomrule
\end{tabular}
\vspace{-0.85em}
\end{table}
In standard decoding across 4K--16K contexts, \sys{} achieves \revisionNov{$1.30\text{--}2.36\times$} higher all-GPU throughput than single-node \texttt{EP4} (Figure~\ref{fig:overview}a, Table~\ref{tab:std}). However, scaling colocated vLLM across the rack boosts baseline throughput by \revisionNov{$22\text{--}114\%$}, peaking at interior widths \texttt{EP16}--\texttt{EP32} rather than \texttt{EP64} on every workload (Figure~\ref{fig:overview}c) and compressing \sys{}'s per-GPU batch advantage to $1.06\text{--}1.16\times$ on the full-attention models (Figure~\ref{fig:overview}b). Against the best wide-EP configuration at each context, \sys{}'s all-GPU speedup $S_{\allgpu}$ narrows to \revisionNov{$0.99\text{--}1.17\times$} (Figure~\ref{fig:overview}a; Appendix~A). Specifically, \mm{} sits at parity with $1.00\text{--}1.02\times$ speedup, whereas \qwen{} retains $1.06\text{--}1.17\times$ all-GPU gains \revisionNov{and \hybrid{} falls from $1.16\times$ at 4K to parity ($0.99\times$) at 16K}. Before charging dedicated FFN GPUs, however, attention-GPU speedup $S_{\attgpu}$ reaches \revisionNov{$1.08\text{--}1.39\times$} across all workloads.

Under speculative decoding at $d=3$ with acceptance controlled near $\bar{a}\approx 1.8$ tokens per step, multi-token verification accelerates both engines across 4K--16K contexts (Figure~\ref{fig:spec}a; Appendix~D). At the same time, verifying four tokens per step contracts \sys{}'s optimal A:F node ratio by more than half on every workload, from 9:1--17:1 down to 4:1--8:1 on \mm{} and from 5:1--11:1 down to 2:1--5:1 on \qwen{} (Figure~\ref{fig:spec}c, Table~\ref{tab:spec}). Despite dedicating a larger GPU share to FFN nodes, \sys{} lifts $S_{\allgpu}$ on \mm{} from standard-decoding parity up to $1.08\times$ while holding \qwen{} at $1.10\text{--}1.13\times$, and raises $S_{\attgpu}$ across all workloads to $1.22\text{--}1.70\times$ (Figure~\ref{fig:spec}b).

\section{Understanding the Performance}
\label{sec:model}
\label{sec:spec}

\begin{figure*}[t]
\centering
\revisionNovFig\includegraphics[width=\textwidth]{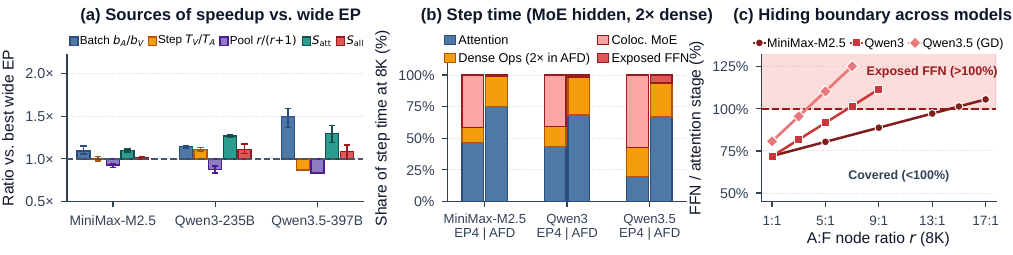}
\vspace{-1.15em}
\caption{\textbf{Three-factor throughput decomposition, step breakdown, and coverage boundary.} \textbf{(a)} Equation~\ref{eq:identity} factors across 4K--16K standard decoding: batch gain $b_A/b_V$, overlap gain $T_V/T_A$, disaggregation cost $r/(r{+}1)$, attention-GPU speedup $S_{\attgpu}$, and all-GPU speedup $S_{\allgpu}$. \textbf{(b)} Normalized 8K step share: \sys{} hides colocated MoE while running attention-stage projections twice per step. \textbf{(c)} Ratio of remote FFN latency to the attention stage at 8K, crossing $100\%$ at each model's measured peak (15:1, 7:1, and 5:1).}
\vspace{-0.85em}
\label{fig:decomposition}
\end{figure*}

\subsection{Speedup Decomposition}
\label{sec:model-decomp}
\label{sec:decomp}

\sys{}'s all-GPU speedup over colocated wide EP is the product of two gains and one cost (Eq.~\ref{eq:identity}). Let $r = N_A/N_F$ be the number of attention nodes per FFN node, $b_A$ and $b_V$ the per-GPU decode batches, and $T_A$ and $T_V$ the decode step times of \sys{} and colocated wide EP. Per-GPU throughput is batch over step time, and only $r$ of every $r{+}1$ nodes emit tokens, so
\begin{equation}
S_{\allgpu} \;=\; \frac{b_A}{b_V}\,\times\,\frac{T_V}{T_A}\,\times\,\frac{r}{r+1}.
\label{eq:decomp}
\end{equation}
The batch gain $b_A/b_V$ comes from removing expert weights from attention GPUs, the overlap gain $T_V/T_A$ from taking the MoE off the critical path, and the disaggregation cost $r/(r{+}1)$ from FFN GPUs that hold no KV cache and serve no requests. The two gains together form the attention-GPU speedup of Section~\ref{sec:setup}, $S_{\attgpu} = (b_A/b_V)(T_V/T_A)$, so $S_{\allgpu} = \frac{r}{r+1}\,S_{\attgpu}$: the disaggregation cost is exactly the gap between what each attention GPU gains and what the rack gains. Equation~\ref{eq:decomp} holds in both standard and speculative decoding when the two engines reach the same mean acceptance.

Figure~\ref{fig:decomposition}a shows how each term plays out on the three models.

\textbf{Batch gain.} Removing expert weights from the attention GPUs helps only where wide EP has not already freed that memory. On \mm{} and \qwen{}, wide EP has done most of the work, so the batch grows only modestly. \hybrid{} is the exception: its batch \revisionNov{grows by $1.4\text{--}1.6\times$}, because a \sys{} attention GPU keeps more HBM for KV cache and linear-attention state than a vLLM GPU, which also holds expert weights and reserves more memory for its runtime, and stores each request more compactly than vLLM's shared hybrid-model pages.

\textbf{Overlap gain.} Taking the MoE off the critical path pays only when the MoE is heavy. \qwen{}, with heavy experts in all 94 layers, gains the most. \mm{}'s light experts leave little to hide, so its step barely changes. \hybrid{}'s step is even slower than wide EP's, because its attention-side projections run twice per step and its short linear-attention layers cannot always hide the FFN.

\textbf{Disaggregation cost.} The cost shrinks as one FFN node serves more attention nodes. On the full-attention models, longer contexts allow this, so their advantage grows with context. \hybrid{}'s linear-attention layers keep its ratio low at every context, so its cost stays high, and its advantage fades as its gains shrink.

Together, the three terms give each model its regime. \mm{} sits near parity because neither gain repays even a small cost. \qwen{} has a batch gain similar to \mm{}'s but adds a real overlap gain, which is enough to win. \hybrid{} \revisionNov{leads on batch alone, despite a slower step, until its advantage shrinks to parity at 16K}.

\subsection{Per-Layer Overlap Model}
\label{sec:model-overlap}
While HBM capacity fixes the batch gain, the overlap gain depends on how two-microbatch pipelining overlaps the attention stage and the FFN stage across the $L$ model layers. Splitting each attention GPU's batch into two ping-pong microbatches overlaps one microbatch's attention stage $T_{a,\ell}$ in layer $\ell$ with the peer microbatch's FFN stage $T_{e,\ell}(r)$ across all $r$ attention nodes, giving the predicted step time
\begin{equation}
 \widehat{T}_A(r) = T_{\mathrm{draft}} + 2\sum_{\ell=1}^{L} \max\bigl\{T_{a,\ell},\; T_{e,\ell}(r)\bigr\}.
\label{eq:ordinary}
\end{equation}
Here $T_{\mathrm{draft}}$ is the draft-and-accept time outside the layer loop, zero in standard decoding; the per-layer terms take the same form in both modes (closed-form hardware rooflines in Appendix~C). In the attention stage, $T_{a,\ell}$ combines core attention KV-cache reads with layer norms, QKV/O projections, and top-$K$ routing on the attention GPU, which run once per microbatch and therefore twice per step even when the FFN stage is hidden (Figure~\ref{fig:decomposition}b). In the FFN stage, $T_{e,\ell}(r)$ combines per-layer NVLink dispatch and expert-weight reads with grouped-GEMM compute over the routed tokens aggregated from $r$ attention nodes, growing monotonically with $r$. Under speculative decoding at draft depth $d$, verifying $q = d + 1$ candidate tokens per request reuses the same prefix KV-cache read on attention GPUs---leaving $T_{a,\ell}$ nearly unchanged---while routing $q$ times as many tokens to the FFN pool and steepening $T_{e,\ell}(r)$. Taking the maximum per layer before summing is essential on hybrid models such as \hybrid{}, where idle slack on full-attention layers cannot cancel FFN exposure on sequential linear-attention layers.

\subsection{Coverage Boundary and Win Condition}
\label{sec:model-boundary}
In Equation~\ref{eq:ordinary}, the FFN stage in layer $\ell$ stays hidden behind the peer microbatch's attention stage whenever $T_{e,\ell}(r) \le T_{a,\ell}$. Below this coverage boundary, remote MoE adds no exposed latency to the critical path, so adding attention nodes per FFN node reduces the disaggregation cost without stretching the step; once the FFN stage exceeds the attention stage (crossing $100\%$ in Figure~\ref{fig:decomposition}c), uncovered FFN execution lengthens the step directly. Combining this overlap bound with Equation~\ref{eq:identity} defines the win condition for \sys{} to beat colocated wide EP on all-GPU throughput:
\begin{equation}
 \frac{b_A}{b_V} \cdot \frac{T_V}{T_A(r)} > 1 + \frac{1}{r}, \quad \text{or}\quad \frac{T_V}{T_A(r)} > 1 + \frac{1}{r}\;\;\text{for } b_A \approx b_V.
\label{eq:boundary}
\end{equation}
In the general form, the batch gain and overlap gain together must outweigh the disaggregation cost. When rack-scale wide EP already reclaims most of the KV-cache memory gap so per-GPU batches nearly match, the second form applies: the overlap gain alone must exceed the disaggregation cost, requiring the colocated MoE time removed from the critical path to pay for both the duplicated second pass of attention-stage projections and any uncovered FFN time.

Together, the coverage boundary and the two forms of Equation~\ref{eq:boundary} explain the behavior across models and decoding modes (Section~\ref{sec:provisioning} analyzes the resulting provisioning trajectories). In standard decoding, longer contexts shrink the per-GPU batch and reduce the tokens routed per attention node while KV-cache scans keep the attention stage long on full-attention models, allowing one FFN node to cover progressively more attention nodes. Because wide EP already narrows their batch gap, these full-attention models follow the matched-batch condition: heavy-expert \qwen{} achieves a large enough overlap gain to clear the disaggregation cost, whereas light-expert \mm{} barely breaks even. On hybrid \hybrid{}, by contrast, linear-attention layers shrink their attention stage alongside the batch and cap the covered ratio early; although duplicated projections make its step slower than wide EP, its \revisionNov{$1.4\text{--}1.6\times$ larger batch satisfies the general form of Equation~\ref{eq:boundary} through batch gain at 4K and 8K and reaches only parity at 16K}. Under speculative decoding~\cite{leviathan,chen,specinfer,medusa,eagle,sdillusion}, verifying multiple draft tokens per request reuses prefix KV reads on attention GPUs while multiplying routed tokens on FFN GPUs, so the FFN stage saturates the attention stage at less than half the node ratio (Figure~\ref{fig:spec}c). At the same time, hiding this multiplied expert work behind the attention stage amplifies the overlap gain, lifting light-expert \mm{} from parity to a clear all-GPU win at 4K and 16K while trading a higher disaggregation cost for a much larger attention-GPU speedup on heavy-expert \qwen{}.

\begin{figure*}[t]
\centering
\includegraphics[width=0.97\textwidth]{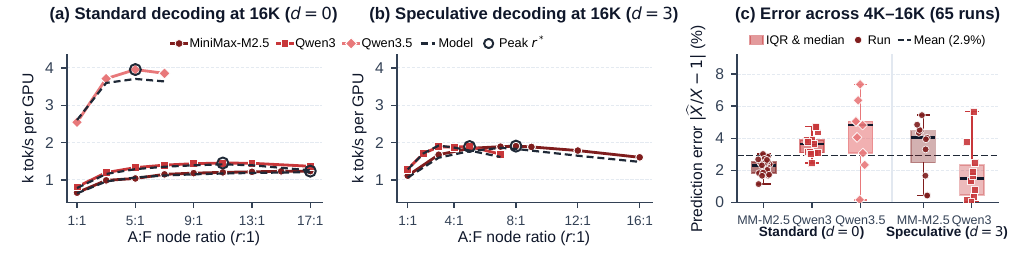}
\vspace{-1.25em}
\caption{\textbf{Zero-refit model validation across candidate A:F ratios.} \textbf{(a)} At 16K standard decoding ($d=0$), predicted throughput (dashed) against measured throughput (solid) across candidate ratios. \textbf{(b)} At 16K speculative decoding ($d=3$), the model captures the leftward shift of the optimal ratio. \textbf{(c)} Relative error $|\widehat X/X-1|$ across all 65 deployments up to each workload's optimal ratio ($2.9\%$ mean error, dashed line).}
\vspace{-0.95em}
\label{fig:validation}
\end{figure*}

\subsection{Model Validation and Ratio Selection}
\label{sec:validation}
We validate Equation~\ref{eq:ordinary} against all 88 measured \sys{} deployments across standard and speculative decoding (complete residuals in Appendix~B). All model inputs come from hardware specifications and a single colocated wide-EP profile per workload, with zero parameters fitted to \sys{} runs (Appendix~C). Across all 65 deployments up to each workload's optimal ratio, the model predicts \sys{} throughput with a mean error of $2.9\%$---specifically $3.0\%$ in standard decoding and $2.6\%$ in speculative decoding---and never exceeds $7.4\%$ on any run (Figure~\ref{fig:validation}c). At 16K (Figures~\ref{fig:validation}a and~\ref{fig:validation}b), the model accurately reproduces both the standard-decoding throughput curves across all three architectures and the leftward shift of the optimal ratio under speculative decoding.

Because the model predicts throughput across candidate ratios without running \sys{}, an operator can pick the deployment ratio directly from Equation~\ref{eq:ordinary} instead of running a multi-node grid search. Deploying the model-selected ratio retains $95\text{--}100\%$ of the measured peak throughput in standard decoding and $97\text{--}100\%$ in speculative decoding, matching the exact measured optimum on 7 of 9 standard workloads and landing one step away on the rest. The coverage condition $T_{e,\ell}(r) \le T_{a,\ell}$ explains why one-step misses at high ratios are benign: while the ratio stays within the coverage bound, step latency remains flat and the disaggregation cost barely changes between adjacent ratios such as $13\text{:}1$, $15\text{:}1$, and $17\text{:}1$. At low ratios or under speculative decoding where fourfold expert load sharpens the peak, testing the model-selected ratio and the candidates within two steps of it finds the exact optimum.

\section{System Implementation and Ablation}
\label{sec:system}

\subsection{System Implementation}
\label{sec:system-impl}
To achieve the theoretical overlap of Equation~\ref{eq:ordinary} without software or communication bubbles, \sys{} co-designs its operator boundary and remote expert kernel around GB200 NVL72's high HBM3e and NVLink bandwidth. In an $r\text{:}1$ deployment, the attention pool comprises $r$ four-GPU nodes running data-parallel attention ranks, and the FFN pool comprises one four-GPU node running expert parallelism across all routed experts. \sys{} instead adapts \texttt{DeepGEMM} into a split \texttt{MegaMoE} architecture: attention GPUs dispatch top-$K$ indices and token counts, while each FFN GPU runs a single persistent \texttt{MegaMoE} kernel per layer that pulls routed FP8 activations over NVLink TMA, executes block-FP8 grouped GEMMs and SwiGLU activation, and pushes BF16 outputs back without host synchronization.

\sys{} pairs this fused kernel with two-level overlap within and across decode steps. Within a single microbatch, each layer's post-MoE residual and the next layer's QKV projection depend sequentially on the expert output returning from the FFN pool. To overlap the two pools, \sys{} splits each attention GPU's decode batch into $k = 2$ ping-pong microbatches captured inside a single whole-step CUDA graph: while the FFN pool executes one microbatch's \texttt{MegaMoE} kernel, the attention pool concurrently executes the peer microbatch's attention stage (attention and projections) (Figure~\ref{fig:arch}b). Across decode steps, a $\sim 30\,\text{ms}$ step would stall at step boundaries if GPUs waited synchronously for the CPU to process sampled tokens and update paged KV-cache tables. \sys{} eliminates this inter-step gap with an asynchronous one-step-ahead coordinator: while GPUs execute the current step's CUDA graph, the CPU processes the previous step's sampled tokens, prepares the next step's KV-cache slots and metadata descriptors, and enqueues the next launch before the current step finishes on the device.

\subsection{Ablation Study}
\label{sec:system-ablation}
\begin{figure}[t]
\centering
\includegraphics[width=0.95\columnwidth]{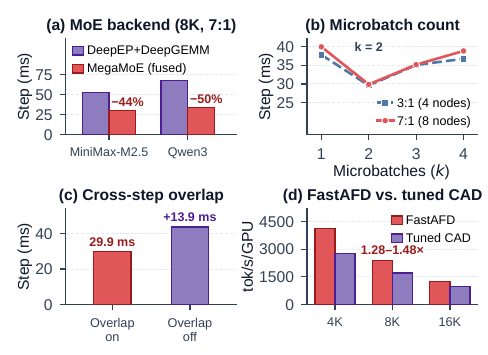}
\vspace{-1.25em}
\caption{\textbf{Runtime ablations at 8K and MoE-boundary vs.\ tuned CAD across 4K--16K.} \textbf{(a)} Fused \texttt{MegaMoE} vs.\ \texttt{DeepEP}+\texttt{DeepGEMM} at 7:1. \textbf{(b)} Microbatch count $k$ on \mm{} at 3:1 and 7:1. \textbf{(c)} One-step-ahead coordinator on vs.\ off at 7:1. \textbf{(d)} \sys{} vs.\ peak-tuned CAD on \mm{} across 4K--16K.}
\vspace{-0.85em}
\label{fig:ablation-cad}
\end{figure}
\begin{figure*}[t]
\centering
\revisionNovFig\includegraphics[width=0.98\textwidth]{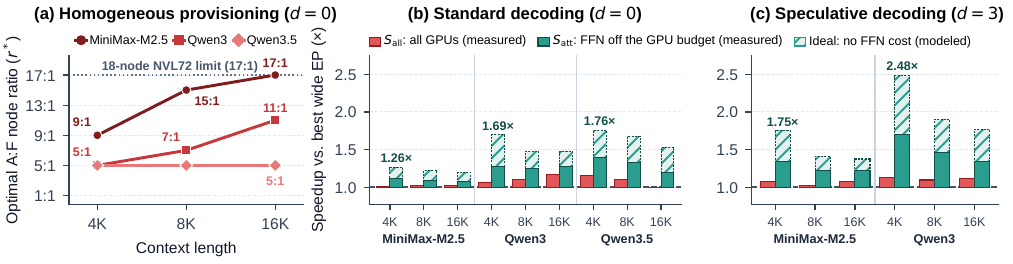}
\vspace{-1.15em}
\caption{\textbf{Rack-scale A:F provisioning and heterogeneous FFN placement across 4K--16K.} \textbf{(a)} Optimal homogeneous node ratios $r^* = N_A/N_F$ in standard decoding ($d=0$): full-attention MoEs fan out from $9{:}1\rightarrow 17{:}1$ (\mm{}) and $5{:}1\rightarrow 11{:}1$ (\qwen{}), whereas hybrid \hybrid{} holds $5{:}1$ across 4K--16K. \textbf{(b, c)} Measured speedup over best wide EP in standard ($d=0$) and speculative ($d=3$) decoding when counting all GPUs ($S_{\allgpu}$, coral) or attention GPUs only ($S_{\attgpu}$, solid teal; up to \revisionNov{$1.39\times$} and $1.70\times$), plus the modeled zero-FFN-cost ceiling (hatched; up to \revisionNov{$1.76\times$} and $2.48\times$).}
\vspace{-0.75em}
\label{fig:hetero}
\end{figure*}
Figure~\ref{fig:ablation-cad} evaluates the contribution of each runtime mechanism at 8K context (complete wall-clock measurements in Appendix~E). First, replacing modular \texttt{DeepEP}+\texttt{DeepGEMM} with the fused \texttt{MegaMoE} kernel cuts step latency by $50\%$ on \qwen{} and by $44\%$ on \mm{} at a $7\text{:}1$ ratio (Figure~\ref{fig:ablation-cad}a), while avoiding over $12\,\text{GiB}$ of extra buffer memory per attention GPU that causes \texttt{DeepEP}+\texttt{DeepGEMM} to run out of memory at full batch on \qwen{}. Second, sweeping the microbatch count from $k = 1$ to $k = 4$ on \mm{} confirms that $k = 2$ achieves the lowest step latency at both $3\text{:}1$ and $7\text{:}1$ (Figure~\ref{fig:ablation-cad}b): $k = 1$ serializes attention and FFN execution, whereas $k = 3$ and $k = 4$ repeat the attention GPUs' dense projections $k$ times per step and shrink per-microbatch GEMM dimensions. Third, disabling the one-step-ahead coordinator at $7\text{:}1$ adds $14\,\text{ms}$ of synchronous host overhead per step, reducing end-to-end throughput by $32\%$ (Figure~\ref{fig:ablation-cad}c).

Finally, we compare \sys{}'s MoE-boundary split against an inference variant of Core Attention Disaggregation (CAD), which DistCA proposed for long-context training~\cite{distca}; the variant moves QKV and output projections to the FFN pool while keeping only core attention and KV-cache writes on the attention GPUs (Appendix~E). On HBM3e, bare core attention is too brief to cover remote execution, and moving the projections to the FFN pool places an NVLink round trip on the attention critical path while burdening the shared FFN node with dense projections from all $r$ attention nodes. Across peak-tuned deployments on \mm{} from 4K to 16K (Figure~\ref{fig:ablation-cad}d), CAD saturates its FFN node early at $2\text{:}1\text{--}4\text{:}1$ ratios---far below \sys{}'s $9\text{:}1\text{--}17\text{:}1$---and achieves only $0.68\text{--}0.78\times$ of \sys{}'s per-GPU throughput, giving \sys{} a $1.28\text{--}1.48\times$ advantage.\looseness=-1

\section{Provisioning and Outlook}
\label{sec:provisioning}
\label{sec:hetero}

\subsection{Rack-Scale Provisioning}
\label{sec:provisioning-rack}
Figure~\ref{fig:hetero}a tracks how the optimal Attention-to-FFN node ratio $r^* = N_A / N_F$ evolves across context lengths from 4K to 16K under standard decoding. On full-attention MoEs (\mm{} and \qwen{}), growing KV-cache footprints steadily increase attention time per token while shrinking the per-node HBM batch ceiling, driving $r^*$ upward from $9\text{:}1$ to $17\text{:}1$ on \mm{} and from $5\text{:}1$ to $11\text{:}1$ on \qwen{}. At 8K--16K on these two models, a single FFN node serves $7\text{--}17$ attention nodes---consuming only $6\text{--}13\%$ of the GPU fleet---and sits on a broad plateau where neighboring integer ratios remain within $0.5\%$ of peak throughput, allowing a single static split to cover wide workload shifts. In contrast, the hybrid \hybrid{} model maintains a flat $5\text{:}1$ ratio across 4K--16K: its 45 Gated-Delta linear-attention layers replace KV-cache scans with fixed-size recurrent state updates that scale with batch size rather than sequence length, keeping total attention time nearly constant across 4K--16K and delaying ratio fan-out until contexts well beyond 16K where its 15 full-attention layers eventually dominate HBM capacity and latency.\looseness=-1

Enabling speculative decoding fundamentally shifts rack-scale provisioning toward a larger FFN pool. Because tree verification processes $q = d + 1$ candidate tokens per request while amortizing KV-cache reads across the draft tokens, it multiplies the routed FFN token volume per step by $q$ without proportionally lengthening attention. As predicted by the coverage condition $T_{e,\ell}(r) \le T_{a,\ell}$ in Equation~\ref{eq:ordinary} and confirmed in Figure~\ref{fig:spec}c, enabling speculative decoding ($d=3,\; q=4$) contracts the optimal A:F ratio by more than half across 4K--16K---shifting $r^*$ from $9\text{:}1\text{--}17\text{:}1$ down to $4\text{:}1\text{--}8\text{:}1$ on \mm{} and from $5\text{:}1\text{--}11\text{:}1$ down to $2\text{:}1\text{--}5\text{:}1$ on \qwen{}---so operators must allocate $11\text{--}33\%$ of the rack to FFN nodes rather than $6\text{--}17\%$ across 4K--16K in standard decoding.

\subsection{Heterogeneous FFN Placement}
\label{sec:provisioning-hetero}
Our speedup decomposition in Equation~\ref{eq:identity} points to a natural architectural evolution for next-generation rack-scale systems such as NVIDIA Vera Rubin NVL72 paired with Groq~3 LPX accelerators~\cite{nvidiagroqlpx}: placing the stateless FFN pool on dedicated high-bandwidth SRAM accelerators while keeping stateful attention and KV caches on HBM-rich GPUs. In a homogeneous GPU deployment, every node assigned to the FFN pool imposes a disaggregation cost of $r/(r+1)$ on the GPU budget that holds KV cache and generates tokens. Placing the FFN pool on non-GPU accelerators removes this disaggregation cost, so per-attention-GPU throughput ($S_{\attgpu}$, solid teal bars in Figures~\ref{fig:hetero}b and~\ref{fig:hetero}c) directly becomes the per-GPU system speedup. Even with today's NVLink communication and overlap schedules unchanged, eliminating the disaggregation cost lifts end-to-end GPU speedups from \revisionNov{$0.99\text{--}1.17\times$} to \revisionNov{$1.08\text{--}1.39\times$} under standard decoding (Figure~\ref{fig:hetero}b) and from $1.02\text{--}1.13\times$ to $1.22\text{--}1.70\times$ under speculative decoding (Figure~\ref{fig:hetero}c). Notably, heterogeneous FFN placement yields a much larger jump under speculative decoding precisely because verification requires a heavier FFN pool ($11\text{--}33\%$ of homogeneous GPUs at $d=3$), so moving FFN off the GPU budget recovers a far larger disaggregation cost.\looseness=-1

A second gain arises when dedicated FFN accelerators execute expert weights from high-bandwidth on-chip SRAM, compressing remote FFN latency toward zero. In that regime, attention nodes no longer need two-microbatch ping-pong pipelining to hide remote FFN execution, eliminating both the second pass of dense projections on attention GPUs and any uncovered FFN tail latency (closed-form bound in Appendix~C). Under this ideal zero-FFN-cost ceiling (hatched bars in Figures~\ref{fig:hetero}b and~\ref{fig:hetero}c), per-GPU speedups reach \revisionNov{$1.20\text{--}1.76\times$} in standard decoding and $1.38\text{--}2.48\times$ in speculative decoding across 4K--16K, turning even near-parity \mm{} into a $1.20\text{--}1.26\times$ standard-decoding and $1.38\text{--}1.75\times$ speculative-decoding win. Together, these bounds show that while homogeneous GPU disaggregation is bounded by the disaggregation cost and two-microbatch pipelining overhead, heterogeneous Attention-GPU + FFN-SRAM disaggregation unlocks both halves of that headroom simultaneously.\looseness=-1

\section{Related Work}
\label{sec:discussion}
Disaggregated LLM serving operates at several orthogonal granularities. Request-phase disaggregation and chunked-prefill schedulers~\cite{distserve,splitwise,mooncake,sarathi} separate compute-bound prefill from memory-bound decoding across distinct node pools; \sys{} operates strictly within the decode phase and can be deployed inside the decode pool of a prefill--decode disaggregated cluster. Within the decode phase, attention offloading places KV caches and attention on memory-optimized devices~\cite{lamina} or across a cluster's pooled GPU memory~\cite{infinitellm} while keeping dense and MoE projections together, whereas \revisionNov{CAD}, proposed for long-context training~\cite{distca}, runs only core attention on a separate device pool. In contrast, MoE-boundary AFD systems~\cite{megascale,step3,eaas,xdeepserve,vllmafd} split each layer specifically at the routed expert interface; our ablation in Section~\ref{sec:system} demonstrates that keeping dense projections in the attention stage is essential on HBM3e racks to cover remote FFN execution and avoid extra NVLink round trips. Finally, while recent rack-scale AFD studies evaluate homogeneous and heterogeneous trade-offs using analytical communication rooflines or offline simulators~\cite{afdchallenges,afdledger}, we provide the first end-to-end hardware audit on GB200 NVL72 against context-tuned wide EP, paired with a zero-refit per-layer overlap model validated across 65 multi-node deployments.\looseness=-1

Our analysis also connects heterogeneous FFN placement with speculative decoding. Earlier heterogeneous MoE systems such as Fiddler~\cite{fiddler} and MoE-Lightning~\cite{moelightning} run large MoE models on memory-constrained GPUs by keeping expert weights in CPU memory and coordinating CPU and GPU execution. Emerging heterogeneous racks such as NVIDIA Vera Rubin NVL72 paired with Groq~3 LPX accelerators~\cite{nvidiagroqlpx} instead run latency-sensitive FFN and MoE execution on the accelerators while GPUs keep attention and KV caches, and SQD~\cite{sqd} splits decode by attention type, placing dense attention layers apart from subquadratic attention layers plus FFN. Similarly, prior speculative decoding studies~\cite{leviathan,chen,specinfer,medusa,eagle,sdillusion} analyze multi-token verification under colocated execution, where shared prefix KV reads and multiplied FFN tokens execute on the same GPUs. We show that in a disaggregated rack, verifying $q$ candidate tokens per request directly contracts the coverage boundary $T_{e,\ell}(r) \le T_{a,\ell}$ in Equation~\ref{eq:ordinary} and shifts cluster provisioning toward a heavier FFN pool---which limits homogeneous all-GPU gains on heavy-expert models, but makes heterogeneous FFN placement substantially more rewarding under speculative decoding than under standard decoding.\looseness=-1

\section{Conclusion}
\label{sec:conclusion}
We asked when attention--FFN disaggregation pays on rack-scale NVLink and answered it by measurement. With \sys{}, an AFD engine for GB200 NVL72 that leaves no runtime slack between the measured gap and the architecture, we find that AFD ranges from parity to \revisionNov{$1.17\times$} over per-workload-tuned wide EP with every FFN GPU counted: a provisioning decision, not an architectural win. We trace the outcome to a batch gain, an overlap gain, and a disaggregation cost, and show that a per-layer model calibrated from colocated profiles alone predicts it within 3\%, so an operator can decide before deploying AFD. Speculative decoding and off-GPU expert accelerators are the two levers that tilt the balance toward AFD.

\label{main:end}
\section*{Acknowledgements}
\revisionNov{We thank NVIDIA for supporting our development on GB200 NVL72 and for providing access to the rack used for all experiments in this paper. We thank the Qwen team and MiniMax for open-sourcing \qwen{}, \hybrid{}, and \mm{}, and Yibo Zhu and Yonghao Zhuang for insightful feedback. \sys{} builds on the open-source SGLang, vLLM, DeepEP, and DeepGEMM projects.}
\clearpage
\bibliography{references}
\bibliographystyle{mlsys2025}
\section*{AI Usage Disclosure}
The authors used Claude (via Claude Code) and Gemini (via Antigravity) as assistive tools for code development in portions of the serving runtime and verification and plotting scripts, manuscript polishing, literature search, and author-directed analysis. The tools also supported benchmark execution and logging under fixed, author-defined configurations. The authors retained control over the research questions, experimental design, and interpretation of results, reviewed the text and code, verified references, and recomputed reported results from raw hardware logs. The authors take full responsibility for the work.

\clearpage
\appendix
\setcounter{table}{0}\setcounter{figure}{0}\setcounter{equation}{0}
\renewcommand{\thetable}{A\arabic{table}}\renewcommand{\theHtable}{A\arabic{table}}
\renewcommand{\thefigure}{A\arabic{figure}}\renewcommand{\theHfigure}{A\arabic{figure}}
\renewcommand{\theequation}{A\arabic{equation}}\renewcommand{\theHequation}{A\arabic{equation}}
These appendices give the complete data and model details behind the main paper. Every value is measured on the GB200 NVL72 rack unless it is marked as predicted or modeled. We write \qwen{} for \qwenfull{} and \hybrid{} for \hybridfull{}.
Appendix~\ref{app:ordinary} defines the measurements and lists all standard-decode results (Section~3).
Appendix~\ref{app:model} lists every model candidate with its prediction error (Section~4.4).
Appendix~\ref{app:profiling} defines the model, its calibration, and the notation (Section~4).
Appendix~\ref{app:spec} gives the complete speculative comparisons (Sections~3 and~6).
Appendix~\ref{app:ablation} reports the runtime ablations and the CAD sweep (Section~5).
Appendix~\ref{app:repro} lists the run presets and software settings.

\section{Model Specifications and Complete Standard-Decode Results}
\label{app:ordinary}
Table~\ref{tab:models} summarizes the architectural parameters of the three evaluated FP8 MoE checkpoints. Table~\ref{tab:best} gives the best measured configuration of each engine for the nine standard-decode workloads (4K--16K), and Tables~\ref{tab:ordinary-mm}--\ref{tab:ordinary-q5b-v033} give every measured configuration (rules in Appendix~\ref{app:measure}). Throughput counts all allocated GPUs, including FFN GPUs, unless it is marked as attention-only; dashes mark configurations without a valid run. Widths past each wide-EP peak were also measured: Qwen3 EP64 at 4K falls to 2986 tokens/s/GPU after peaking at EP16. Every comparison uses the best measured width.

\begin{table}[!htbp]
\caption{Evaluated FP8 MoE models and architectural parameters.}
\label{tab:models}\centering\scriptsize
\setlength{\tabcolsep}{2.5pt}
\begin{tabular}{@{}llrrrr@{}}\toprule
Model & Attention & Total / Active & $L$ & Experts & Top-$K$\\\midrule
\mm{} & Full (GQA) & 229B / 10B & 62 & 256 & 8\\
Qwen3-235B-A22B & Full (GQA) & 235B / 22B & 94 & 128 & 8\\
Qwen3.5-397B-A17B & 15 Full + 45 Linear & 397B / 17B & 60 & 512 & 10\\\bottomrule
\end{tabular}
\end{table}

The throughput tables and the model-error tables (Appendix~\ref{app:model}) come from the same runs and can differ in the last digit through rounding or the choice among repeated readings.
\begin{table*}[!htbp]
\centering\small
\caption{Best measured standard-decode configuration of each engine per workload (tokens/s/GPU, all GPUs counted). Optima are over the measured A:F ratios and EP widths.}\label{tab:best}
\begin{tabular}{lllr lrrr}\toprule
Model & Context & AFD & Throughput & Baseline & Throughput & / EP4 & / best EP\\\midrule
MiniMax-M2.5 & 4K & AFD 9:1 & 4118.0 & vLLM EP32 & 4098.0 & 1.403 & 1.005\\
MiniMax-M2.5 & 8K & AFD 15:1 & 2372.0 & vLLM EP32 & 2317.0 & 1.518 & 1.024\\
MiniMax-M2.5 & 16K & AFD 17:1 & 1234.0 & vLLM EP32 & 1208.0 & 1.556 & 1.022\\
Qwen3 & 4K & AFD 5:1 & 4256.0 & vLLM EP16 & 4004.0 & 1.302 & 1.063\\
Qwen3 & 8K & AFD 7:1 & 2624.0 & vLLM EP16 & 2389.0 & 1.412 & 1.098\\
Qwen3 & 16K & AFD 11:1 & 1452.0 & vLLM EP32 & 1240.0 & 1.425 & 1.171\\
Qwen3.5 & 4K & AFD 5:1 & 6968.0 & vLLM EP32 & \revisionNov{6012.0} & \revisionNov{2.254} & \revisionNov{1.159}\\
Qwen3.5 & 8K & AFD 5:1 & 5627.0 & vLLM EP32 & \revisionNov{5095.0} & \revisionNov{2.362} & \revisionNov{1.104}\\
Qwen3.5 & 16K & AFD 5:1 & 3955.0 & vLLM EP32 & \revisionNov{3982.0} & \revisionNov{2.114} & \revisionNov{0.993}\\
\bottomrule\end{tabular}
\end{table*}

\begin{table}[!htbp]
\centering\small
\caption{MiniMax-M2.5 standard-decode throughput (tokens/s/GPU, counting FFN GPUs). Dashes mark configurations without a valid run.}\label{tab:ordinary-mm}
\begin{tabular}{lrrr}\toprule
Configuration & 4K & 8K & 16K\\\midrule
AFD 1:1 & 2341.0 & 1281.0 & 657.0\\
AFD 3:1 & 3500.0 & 1920.0 & 987.0\\
AFD 5:1 & 3883.0 & 2128.0 & 1038.0\\
AFD 7:1 & 4042.0 & 2230.0 & 1148.0\\
AFD 9:1 & 4118.0 & 2294.0 & 1180.0\\
AFD 11:1 & 3553.0 & 2331.0 & 1201.0\\
AFD 13:1 & --- & 2364.0 & 1216.0\\
AFD 15:1 & --- & 2372.0 & 1229.0\\
AFD 17:1 & --- & 2249.0 & 1234.0\\
vLLM EP4 & 2936.0 & 1563.0 & 793.0\\
vLLM EP8 & 3664.0 & 1942.0 & 1028.0\\
vLLM EP16 & 4076.0 & 2251.0 & 1159.0\\
vLLM EP32 & 4098.0 & 2317.0 & 1208.0\\
vLLM EP64 & 3812.0 & 2029.0 & 1119.0\\
\bottomrule\end{tabular}
\end{table}

\begin{table}[!htbp]
\centering\small
\caption{Qwen3 standard-decode throughput (tokens/s/GPU, counting FFN GPUs). Dashes mark configurations without a valid run.}\label{tab:ordinary-qw}
\begin{tabular}{lrrr}\toprule
Configuration & 4K & 8K & 16K\\\midrule
AFD 1:1 & 2721.0 & 1542.0 & 810.0\\
AFD 3:1 & 4055.0 & 2289.0 & 1195.0\\
AFD 5:1 & 4256.0 & 2542.0 & 1329.0\\
AFD 7:1 & 3907.0 & 2624.0 & 1396.0\\
AFD 9:1 & --- & 2530.0 & 1434.0\\
AFD 11:1 & --- & --- & 1452.0\\
AFD 13:1 & --- & --- & 1444.0\\
AFD 17:1 & --- & --- & 1365.0\\
vLLM EP4 & 3270.0 & 1859.0 & 1019.0\\
vLLM EP8 & 3795.0 & 2173.0 & 1157.0\\
vLLM EP16 & 4004.0 & 2389.0 & 1227.0\\
vLLM EP32 & 3728.0 & 2310.0 & 1240.0\\
vLLM EP64 & 2986.0 & 2073.0 & 1148.0\\
\bottomrule\end{tabular}
\end{table}

\begin{table}[!htbp]
\centering\small
\caption{Qwen3.5 standard-decode throughput (tokens/s/GPU, counting FFN GPUs). Dashes mark configurations without a valid run.}\label{tab:ordinary-q5b-v033}
\begin{tabular}{lrrr}\toprule
Configuration & 4K & 8K & 16K\\\midrule
AFD 1:1 & 4366.0 & 3532.0 & 2543.0\\
AFD 3:1 & 6494.0 & 5220.0 & 3711.0\\
AFD 5:1 & 6968.0 & 5627.0 & 3955.0\\
AFD 7:1 & 4716.0 & 5054.0 & 3854.0\\
vLLM EP4 & \revisionNov{3092.0} & \revisionNov{2382.0} & \revisionNov{1871.0}\\
vLLM EP8 & \revisionNov{5186.0} & \revisionNov{4188.0} & \revisionNov{3226.0}\\
vLLM EP16 & \revisionNov{5950.0} & \revisionNov{5037.0} & \revisionNov{3762.0}\\
vLLM EP32 & \revisionNov{6012.0} & \revisionNov{5095.0} & \revisionNov{3982.0}\\
vLLM EP64 & \revisionNov{5605.0} & \revisionNov{4895.0} & \revisionNov{3770.0}\\
\bottomrule\end{tabular}
\end{table}

\subsection{How Each Number Is Measured}
\label{app:measure}
\paragraph{\sys{} cells.} Each request decodes exactly 1024 output tokens at a fixed per-attention-GPU batch $b_A$: the KV-full batch, or a slightly smaller batch aligned to whole attention-kernel waves when that runs faster. After the last prefill step, the period $T_A$ is the mean decode step over the following steady decode window, and throughput counts every allocated GPU, $X_{\allgpu}=\frac{r}{r+1}\,b_A/T_A$. Cells with several readings report their mean.

\paragraph{vLLM cells.} Every GPU runs one engine (TP1; EP equals the number of GPUs), and each engine holds the KV-full batch $b_V=\lfloor \text{KV tokens per GPU}/(\text{ISL}+\text{OSL})\rfloor$, stepping down when a batch yields no valid window. Eight streaming probe requests timestamp every output token on the client; because every step is an all-to-all over all engines, a probe's inter-token interval is the global step time. The window starts at the first metrics sample where all requests are running and the prompt-token counter has reached its total, and ends at the first completion; $T_V$ is the pooled mean of the probe intervals inside it, and throughput is $b_V/T_V$ per GPU. A valid window needs at least 200 intervals. When the strict window is shorter or never starts (the prompt counter can stop just below its total), a fallback window is accepted only if it starts after prefill with margin, ends before the first completion, contains no prefill, queueing, or preemption, and its largest interval is at most $6.85\times$ the median; probes that deliver tokens in bursts (over 1.15 tokens per interval) are excluded. Each request decodes $\text{OSL}=1024$ output tokens (Table~\ref{tab:presets}).

\paragraph{Ablations.} Ablation periods (Table~\ref{tab:ablation}) are wall-clock: the inverse slope of the decode-step index against log timestamps over the window, so host gaps between steps count.

\section{All Model Candidates and Residuals}
\label{app:model}
Tables~\ref{tab:ordinary-errors-1}--\ref{tab:speculative-errors-2} list every measured candidate, 53 standard and 76 speculative ($d\in\{3,7\}$), with its predicted period. Signed errors are $e=100(T/\widehat T-1)$, which equals the relative throughput error at fixed batch and useful-token output. ``Scope'' marks ratios at or below the measured peak of the same model, context, and draft length; candidates past the peak are kept. Table~\ref{tab:validation-4k16k} summarizes accuracy and zero-sweep ratio selection across the 88 primary 4K--16K runs evaluated in Section~4.4, and Table~\ref{tab:error-summary} extends the error summary to 32K and $d=7$. Selection loss compares the measured throughput at the predicted best ratio with the measured best of the same workload; candidate ratios are discrete and bounded by the 18-node rack.

\begin{table}[!htbp]
\caption{Model accuracy and ratio-choice summary against the 88 measured \sys{} runs at 4K--16K (Section~4.4). Error is $|\widehat X/X-1|$.}
\label{tab:validation-4k16k}\centering\footnotesize
\setlength{\tabcolsep}{3pt}
\begin{tabular}{@{}lcc@{}}\toprule
 & Standard & Speculative\\
 & (4K--16K) & ($d=3$, 4K--16K)\\\midrule
Workloads & 9 & 6\\
Mean error at the peak & 3.41\% & 4.07\%\\
Mean error & 3.04\% & 2.64\%\\
Max error at the peak & 7.36\% & 5.66\%\\
Max error & 7.36\% & 5.66\%\\
Model picks the best ratio & 7 of 9 & 2 of 6\\
Peak throughput kept by its pick & 95.3--100\% & 96.5--100\%\\
\bottomrule
\multicolumn{3}{@{}p{0.97\columnwidth}@{}}{\rule{0pt}{2.6ex}\scriptsize Mean and max error cover every A:F ratio up to each workload's measured best ratio $r^*$ (44 standard and 21 speculative runs); ``at the peak'' is the run at $r^*$ alone. Past the peak (9 and 14 runs), errors average 10.2\% and 6.6\%. ``Its pick'' is the ratio with the highest predicted throughput.}\\

\end{tabular}
\end{table}

\begin{table}[!htbp]
\caption{Extended model error summary including 32K and $d=7$: mean and maximum $|e|$ (\%) over the listed runs, and the worst selection loss.}
\label{tab:error-summary}\centering\footnotesize
\setlength{\tabcolsep}{3pt}
\begin{tabular}{@{}llrrrr@{}}\toprule
Decoding & Runs & $n$ & Mean & Max & Sel.\ loss\\\midrule
Standard ($d=0$) & all & 53 & 4.25 & 43.38 & 4.72\\
 & $r\le r^*$ & 44 & 3.04 & 7.36 & \\
 & peaks & 9 & 3.41 & 7.36 & \\\midrule
$d=3$ & all & 42 & 4.32 & 11.05 & 3.49\\
 & $r\le r^*$ & 26 & 3.19 & 8.77 & \\
 & peaks & 8 & 4.32 & 8.30 & \\\midrule
$d\in\{3,7\}$ & all & 76 & 5.26 & 16.81 & 10.90\\
 & $r\le r^*$ & 50 & 4.91 & 16.81 & \\
 & peaks & 16 & 4.22 & 8.30 & \\\bottomrule
\end{tabular}
\end{table}

In standard decoding at $r\le r^*$, the median error is 2.69\% and the IQR is $2.24\text{--}3.80\%$. Under $d=3$ at $r\le r^*$, the mean is 2.64\% with a 5.66\% maximum across 4K--16K, and the maximum is 8.77\% at 32K. The largest standard error (43.38\%) is Qwen3.5 4K at 7:1, past the peak; the largest speculative error (16.81\%) is Qwen3 16K at $d=7$ and 2:1, within scope.
\begin{table}[!htbp]
\centering\scriptsize\setlength{\tabcolsep}{2.6pt}
\caption{Standard-decode model candidates 1--27: measured period $T$, predicted period $\widehat T$, and signed error $e=100(T/\widehat T-1)$. Scope Y marks $r\le r^*$.}\label{tab:ordinary-errors-1}
\begin{tabular}{@{}llrrrrrrc@{}}\toprule
Model & Context & $d$ & A:F & $b_A$ & $T$ (ms) & $\widehat T$ (ms) & $e$ (\%) & Scope\\\midrule
MiniMax-M2.5 & 4K & 0 & 1:1 & 156 & 33.320 & 34.047 & -2.13 & Y\\
MiniMax-M2.5 & 4K & 0 & 3:1 & 156 & 33.420 & 34.047 & -1.84 & Y\\
MiniMax-M2.5 & 4K & 0 & 5:1 & 156 & 33.480 & 34.047 & -1.66 & Y\\
MiniMax-M2.5 & 4K & 0 & 7:1 & 154 & 33.340 & 34.047 & -2.08 & Y\\
MiniMax-M2.5 & 4K & 0 & 9:1 & 154 & 33.660 & 34.047 & -1.14 & Y\\
MiniMax-M2.5 & 4K & 0 & 11:1 & 152 & 39.210 & 35.257 & +11.21 & N\\
MiniMax-M2.5 & 8K & 0 & 1:1 & 76 & 29.660 & 30.382 & -2.38 & Y\\
MiniMax-M2.5 & 8K & 0 & 3:1 & 76 & 29.690 & 30.382 & -2.28 & Y\\
MiniMax-M2.5 & 8K & 0 & 5:1 & 76 & 29.770 & 30.382 & -2.02 & Y\\
MiniMax-M2.5 & 8K & 0 & 7:1 & 76 & 29.830 & 30.382 & -1.82 & Y\\
MiniMax-M2.5 & 8K & 0 & 9:1 & 76 & 29.820 & 30.382 & -1.85 & Y\\
MiniMax-M2.5 & 8K & 0 & 11:1 & 76 & 29.880 & 30.382 & -1.65 & Y\\
MiniMax-M2.5 & 8K & 0 & 13:1 & 76 & 29.860 & 30.382 & -1.72 & Y\\
MiniMax-M2.5 & 8K & 0 & 15:1 & 76 & 30.040 & 30.794 & -2.45 & Y\\
MiniMax-M2.5 & 8K & 0 & 17:1 & 76 & 31.920 & 32.069 & -0.46 & N\\
MiniMax-M2.5 & 16K & 0 & 1:1 & 38 & 28.920 & 29.766 & -2.84 & Y\\
MiniMax-M2.5 & 16K & 0 & 3:1 & 38 & 28.870 & 29.766 & -3.01 & Y\\
MiniMax-M2.5 & 16K & 0 & 5:1 & 38 & 30.500 & 29.766 & +2.46 & Y\\
MiniMax-M2.5 & 16K & 0 & 7:1 & 38 & 28.960 & 29.766 & -2.71 & Y\\
MiniMax-M2.5 & 16K & 0 & 9:1 & 38 & 28.970 & 29.766 & -2.68 & Y\\
MiniMax-M2.5 & 16K & 0 & 11:1 & 38 & 29.010 & 29.766 & -2.54 & Y\\
MiniMax-M2.5 & 16K & 0 & 13:1 & 38 & 29.020 & 29.766 & -2.51 & Y\\
MiniMax-M2.5 & 16K & 0 & 15:1 & 38 & 28.980 & 29.766 & -2.64 & Y\\
MiniMax-M2.5 & 16K & 0 & 17:1 & 38 & 29.080 & 29.766 & -2.31 & Y\\
Qwen3.5 & 4K & 0 & 1:1 & 527 & 60.350 & 57.457 & +5.03 & Y\\
Qwen3.5 & 4K & 0 & 3:1 & 515 & 59.480 & 56.623 & +5.05 & Y\\
Qwen3.5 & 4K & 0 & 5:1 & 503 & 60.155 & 56.032 & +7.36 & Y\\
\bottomrule\end{tabular}
\end{table}

\begin{table}[!htbp]
\centering\scriptsize\setlength{\tabcolsep}{2.6pt}
\caption{Standard-decode model candidates 28--53: measured period $T$, predicted period $\widehat T$, and signed error $e=100(T/\widehat T-1)$. Scope Y marks $r\le r^*$.}\label{tab:ordinary-errors-2}
\begin{tabular}{@{}llrrrrrrc@{}}\toprule
Model & Context & $d$ & A:F & $b_A$ & $T$ (ms) & $\widehat T$ (ms) & $e$ (\%) & Scope\\\midrule
Qwen3.5 & 4K & 0 & 7:1 & 491 & 91.092 & 63.532 & +43.38 & N\\
Qwen3.5 & 8K & 0 & 1:1 & 383 & 54.220 & 52.110 & +4.05 & Y\\
Qwen3.5 & 8K & 0 & 3:1 & 373 & 53.590 & 51.127 & +4.82 & Y\\
Qwen3.5 & 8K & 0 & 5:1 & 365 & 54.050 & 53.966 & +0.16 & Y\\
Qwen3.5 & 8K & 0 & 7:1 & 357 & 61.804 & 58.526 & +5.60 & N\\
Qwen3.5 & 16K & 0 & 1:1 & 247 & 48.570 & 47.466 & +2.33 & Y\\
Qwen3.5 & 16K & 0 & 3:1 & 241 & 48.710 & 50.252 & -3.07 & Y\\
Qwen3.5 & 16K & 0 & 5:1 & 235 & 49.513 & 52.875 & -6.36 & Y\\
Qwen3.5 & 16K & 0 & 7:1 & 231 & 52.452 & 55.576 & -5.62 & N\\
Qwen3 & 4K & 0 & 1:1 & 202 & 37.120 & 35.557 & +4.40 & Y\\
Qwen3 & 4K & 0 & 3:1 & 200 & 36.990 & 35.320 & +4.73 & Y\\
Qwen3 & 4K & 0 & 5:1 & 200 & 39.160 & 41.102 & -4.72 & Y\\
Qwen3 & 4K & 0 & 7:1 & 198 & 44.340 & 47.649 & -6.94 & N\\
Qwen3 & 8K & 0 & 1:1 & 108 & 35.010 & 36.139 & -3.12 & Y\\
Qwen3 & 8K & 0 & 3:1 & 106 & 34.730 & 36.139 & -3.90 & Y\\
Qwen3 & 8K & 0 & 5:1 & 106 & 34.740 & 36.139 & -3.87 & Y\\
Qwen3 & 8K & 0 & 7:1 & 106 & 35.340 & 36.726 & -3.77 & Y\\
Qwen3 & 8K & 0 & 9:1 & 106 & 37.700 & 40.322 & -6.50 & N\\
Qwen3 & 16K & 0 & 1:1 & 56 & 34.560 & 35.862 & -3.63 & Y\\
Qwen3 & 16K & 0 & 3:1 & 54 & 33.880 & 34.954 & -3.07 & Y\\
Qwen3 & 16K & 0 & 5:1 & 54 & 33.870 & 34.954 & -3.10 & Y\\
Qwen3 & 16K & 0 & 7:1 & 54 & 33.850 & 34.954 & -3.16 & Y\\
Qwen3 & 16K & 0 & 9:1 & 54 & 33.900 & 34.954 & -3.02 & Y\\
Qwen3 & 16K & 0 & 11:1 & 54 & 34.100 & 34.954 & -2.44 & Y\\
Qwen3 & 16K & 0 & 13:1 & 54 & 34.710 & 36.048 & -3.71 & N\\
Qwen3 & 16K & 0 & 17:1 & 52 & 35.990 & 39.135 & -8.04 & N\\
\bottomrule\end{tabular}
\end{table}

\begin{table}[!htbp]
\centering\scriptsize\setlength{\tabcolsep}{2.6pt}
\caption{Speculative model candidates 1--38: measured period $T$, predicted period $\widehat T$, and signed error $e=100(T/\widehat T-1)$. Scope Y marks $r\le r^*$.}\label{tab:speculative-errors-1}
\begin{tabular}{@{}llrrrrrrc@{}}\toprule
Model & Context & $d$ & A:F & $b_A$ & $T$ (ms) & $\widehat T$ (ms) & $e$ (\%) & Scope\\\midrule
MiniMax-M2.5 & 4K & 3 & 1:1 & 148 & 36.596 & 36.438 & +0.43 & Y\\
MiniMax-M2.5 & 4K & 3 & 3:1 & 150 & 38.874 & 38.711 & +0.42 & Y\\
MiniMax-M2.5 & 4K & 3 & 4:1 & 150 & 41.367 & 43.745 & -5.44 & Y\\
MiniMax-M2.5 & 4K & 3 & 5:1 & 150 & 44.305 & 48.779 & -9.17 & N\\
MiniMax-M2.5 & 4K & 3 & 7:1 & 148 & 54.269 & 58.362 & -7.01 & N\\
MiniMax-M2.5 & 4K & 7 & 1:1 & 150 & 46.196 & 42.815 & +7.90 & Y\\
MiniMax-M2.5 & 4K & 7 & 2:1 & 150 & 48.868 & 46.399 & +5.32 & Y\\
MiniMax-M2.5 & 4K & 7 & 3:1 & 150 & 53.635 & 56.467 & -5.02 & Y\\
MiniMax-M2.5 & 4K & 7 & 4:1 & 148 & 60.772 & 65.963 & -7.87 & N\\
MiniMax-M2.5 & 8K & 3 & 1:1 & 76 & 32.721 & 34.242 & -4.44 & Y\\
MiniMax-M2.5 & 8K & 3 & 3:1 & 76 & 33.112 & 34.242 & -3.30 & Y\\
MiniMax-M2.5 & 8K & 3 & 5:1 & 76 & 34.465 & 36.024 & -4.33 & Y\\
MiniMax-M2.5 & 8K & 3 & 6:1 & 76 & 36.540 & 38.574 & -5.27 & N\\
MiniMax-M2.5 & 8K & 3 & 7:1 & 76 & 38.484 & 41.125 & -6.42 & N\\
MiniMax-M2.5 & 8K & 7 & 2:1 & 76 & 40.052 & 38.462 & +4.13 & Y\\
MiniMax-M2.5 & 8K & 7 & 3:1 & 76 & 40.309 & 40.991 & -1.66 & Y\\
MiniMax-M2.5 & 8K & 7 & 4:1 & 76 & 42.949 & 46.092 & -6.82 & Y\\
MiniMax-M2.5 & 8K & 7 & 5:1 & 76 & 46.374 & 51.194 & -9.41 & N\\
MiniMax-M2.5 & 16K & 3 & 1:1 & 38 & 30.789 & 32.244 & -4.51 & Y\\
MiniMax-M2.5 & 16K & 3 & 3:1 & 38 & 30.683 & 32.244 & -4.84 & Y\\
MiniMax-M2.5 & 16K & 3 & 5:1 & 38 & 30.945 & 32.244 & -4.03 & Y\\
MiniMax-M2.5 & 16K & 3 & 7:1 & 38 & 31.709 & 32.244 & -1.66 & Y\\
MiniMax-M2.5 & 16K & 3 & 8:1 & 38 & 31.973 & 33.282 & -3.93 & Y\\
MiniMax-M2.5 & 16K & 3 & 9:1 & 38 & 32.740 & 34.558 & -5.26 & N\\
MiniMax-M2.5 & 16K & 3 & 12:1 & 38 & 35.412 & 38.383 & -7.74 & N\\
MiniMax-M2.5 & 16K & 3 & 16:1 & 38 & 40.173 & 43.485 & -7.62 & N\\
MiniMax-M2.5 & 16K & 7 & 4:1 & 38 & 40.521 & 35.409 & +14.44 & Y\\
MiniMax-M2.5 & 16K & 7 & 5:1 & 38 & 39.579 & 37.960 & +4.27 & Y\\
MiniMax-M2.5 & 16K & 7 & 6:1 & 38 & 38.683 & 40.510 & -4.51 & Y\\
MiniMax-M2.5 & 16K & 7 & 7:1 & 38 & 39.645 & 43.061 & -7.93 & N\\
MiniMax-M2.5 & 16K & 7 & 8:1 & 38 & 41.937 & 45.612 & -8.06 & N\\
MiniMax-M2.5 & 32K & 3 & 12:1 & 20 & 32.945 & 31.472 & +4.68 & Y\\
MiniMax-M2.5 & 32K & 3 & 13:1 & 20 & 33.027 & 31.762 & +3.98 & Y\\
MiniMax-M2.5 & 32K & 3 & 14:1 & 20 & 33.039 & 32.434 & +1.87 & Y\\
MiniMax-M2.5 & 32K & 3 & 15:1 & 20 & 33.203 & 33.105 & +0.30 & N\\
MiniMax-M2.5 & 32K & 7 & 5:1 & 20 & 37.332 & 34.249 & +9.00 & Y\\
MiniMax-M2.5 & 32K & 7 & 6:1 & 20 & 37.495 & 34.249 & +9.48 & Y\\
MiniMax-M2.5 & 32K & 7 & 7:1 & 20 & 37.440 & 34.492 & +8.55 & Y\\
\bottomrule\end{tabular}
\end{table}

\begin{table}[!htbp]
\centering\scriptsize\setlength{\tabcolsep}{2.6pt}
\caption{Speculative model candidates 39--76: measured period $T$, predicted period $\widehat T$, and signed error $e=100(T/\widehat T-1)$. Scope Y marks $r\le r^*$.}\label{tab:speculative-errors-2}
\begin{tabular}{@{}llrrrrrrc@{}}\toprule
Model & Context & $d$ & A:F & $b_A$ & $T$ (ms) & $\widehat T$ (ms) & $e$ (\%) & Scope\\\midrule
MiniMax-M2.5 & 32K & 7 & 8:1 & 20 & 37.605 & 35.835 & +4.94 & Y\\
MiniMax-M2.5 & 32K & 7 & 9:1 & 20 & 38.077 & 37.177 & +2.42 & N\\
Qwen3 & 4K & 3 & 1:1 & 194 & 46.352 & 45.227 & +2.49 & Y\\
Qwen3 & 4K & 3 & 2:1 & 194 & 52.397 & 53.096 & -1.32 & Y\\
Qwen3 & 4K & 3 & 3:1 & 192 & 62.060 & 65.843 & -5.75 & N\\
Qwen3 & 4K & 3 & 5:1 & 192 & 98.167 & 91.895 & +6.83 & N\\
Qwen3 & 4K & 3 & 7:1 & 190 & 129.910 & 116.989 & +11.05 & N\\
Qwen3 & 4K & 7 & 1:1 & 192 & 64.375 & 55.820 & +15.33 & Y\\
Qwen3 & 4K & 7 & 2:1 & 192 & 85.266 & 81.546 & +4.56 & Y\\
Qwen3 & 4K & 7 & 3:1 & 192 & 116.935 & 107.598 & +8.68 & N\\
Qwen3 & 8K & 3 & 1:1 & 104 & 41.419 & 42.101 & -1.62 & Y\\
Qwen3 & 8K & 3 & 2:1 & 104 & 42.159 & 42.101 & +0.14 & Y\\
Qwen3 & 8K & 3 & 3:1 & 104 & 45.775 & 47.559 & -3.75 & Y\\
Qwen3 & 8K & 3 & 4:1 & 102 & 50.510 & 54.057 & -6.56 & N\\
Qwen3 & 8K & 3 & 5:1 & 102 & 58.734 & 60.976 & -3.68 & N\\
Qwen3 & 8K & 3 & 7:1 & 102 & 79.190 & 74.816 & +5.85 & N\\
Qwen3 & 8K & 7 & 1:1 & 104 & 55.688 & 50.667 & +9.91 & Y\\
Qwen3 & 8K & 7 & 2:1 & 104 & 58.732 & 57.008 & +3.02 & Y\\
Qwen3 & 8K & 7 & 3:1 & 102 & 70.467 & 70.271 & +0.28 & N\\
Qwen3 & 16K & 3 & 1:1 & 54 & 37.778 & 37.907 & -0.34 & Y\\
Qwen3 & 16K & 3 & 2:1 & 54 & 37.887 & 37.907 & -0.05 & Y\\
Qwen3 & 16K & 3 & 3:1 & 54 & 38.633 & 37.907 & +1.91 & Y\\
Qwen3 & 16K & 3 & 4:1 & 52 & 39.781 & 40.099 & -0.79 & Y\\
Qwen3 & 16K & 3 & 5:1 & 52 & 41.157 & 43.627 & -5.66 & Y\\
Qwen3 & 16K & 3 & 7:1 & 52 & 48.428 & 50.683 & -4.45 & N\\
Qwen3 & 16K & 7 & 1:1 & 54 & 49.070 & 43.649 & +12.42 & Y\\
Qwen3 & 16K & 7 & 2:1 & 54 & 50.986 & 43.649 & +16.81 & Y\\
Qwen3 & 16K & 7 & 3:1 & 54 & 52.210 & 50.258 & +3.88 & Y\\
Qwen3 & 16K & 7 & 4:1 & 52 & 53.971 & 56.432 & -4.36 & N\\
Qwen3 & 32K & 3 & 7:1 & 26 & 38.686 & 42.407 & -8.77 & Y\\
Qwen3 & 32K & 3 & 8:1 & 26 & 38.886 & 42.407 & -8.30 & Y\\
Qwen3 & 32K & 3 & 9:1 & 26 & 40.007 & 42.407 & -5.66 & N\\
Qwen3 & 32K & 7 & 3:1 & 26 & 47.365 & 49.722 & -4.74 & Y\\
Qwen3 & 32K & 7 & 4:1 & 26 & 47.725 & 49.722 & -4.02 & Y\\
Qwen3 & 32K & 7 & 5:1 & 26 & 48.857 & 49.722 & -1.74 & Y\\
Qwen3 & 32K & 7 & 6:1 & 26 & 49.840 & 49.722 & +0.24 & Y\\
Qwen3 & 32K & 7 & 7:1 & 26 & 51.209 & 52.694 & -2.82 & N\\
Qwen3 & 32K & 7 & 8:1 & 26 & 54.288 & 56.222 & -3.44 & N\\
\bottomrule\end{tabular}
\end{table}

\clearpage

\section{Model Definitions, Calibration, and Profiling}
\label{app:profiling}
Table~\ref{tab:notation} defines every symbol of the main paper. The rest of this appendix gives the per-layer model, its calibration, and the speculative extension.
\begin{table*}[!t]
\caption{Notation used in the main paper. Source: \emph{config} = chosen setting; \emph{measured} = run records; \emph{profile} = colocated \texttt{vLLM EP4} profiles; \emph{modeled} = this appendix's model.}
\label{tab:notation}\centering\small
\setlength{\tabcolsep}{3pt}
\begin{tabular}{@{}l p{11.2cm} l@{}}\toprule
Symbol & Meaning & Source\\\midrule
\multicolumn{3}{@{}l}{\emph{Configuration and measured quantities}}\\
$N_A, N_F$ & attention / FFN nodes (four GPUs each) & config\\
$r$ & A:F node ratio $N_A/N_F$ & config\\
$b_A, b_V$ & per-GPU decode batch: \sys{} attention GPU / colocated vLLM & measured\\
$n$ & requests per microbatch, $b_A/2$ & config\\
$T_A, T_V$ & decode step period: \sys{} / vLLM & measured\\
$\widehat T_A$ & predicted \sys{} period (Equation~3) & modeled\\
$X_{\allgpu}$ & decode tokens/s per allocated GPU & measured\\
$S_{\mathrm{EP4}}, S_{\allgpu}$ & \sys{} speedup over EP4 / best wide EP, all GPUs & measured\\
$S_{\attgpu}$ & speedup per attention GPU, $\frac{r+1}{r}S_{\allgpu}$ & measured\\
$S^{\star}$ & zero-FFN-cost ceiling, $(b_A/b_V)\,T_V/(\widehat T_A(r_{\min})-D_V)$ & modeled\\
$L, \ell$ & number of layers, layer index & model config\\\midrule
\multicolumn{3}{@{}l}{\emph{Step components}}\\
$A_V, D_V, M_V$ & vLLM attention, auxiliary (projections, norms, router), and MoE (expert GEMMs plus dispatch/combine) time per step & profile\\
$A_\ell(n)$ & attention time of layer $\ell$ for $n$ requests & modeled\\
$A_A(n)$ & $\sum_\ell A_\ell(n)$ & modeled\\
$F_\ell(m)$ & remote FFN service of layer $\ell$ for $m$ routed tokens (Equation~\ref{eq:app-ffn}) & modeled\\
$m$ & routed tokens per layer: $rn$ standard, $rnq$ speculative & config\\
$T_{a,\ell}, T_{e,\ell}(r)$ & per-layer attention stage $A_\ell(n)+D_V/L$ / FFN stage $F_\ell(rnq)$ (Section~4.2) & modeled\\
$T_{\mathrm{exp}}(r)$ & exposed FFN time, $2\sum_\ell\max\{0,F_\ell(rn)-A_\ell(n)-D_V/L\}$; since $\max\{x,y\}=x+\max\{0,y-x\}$, in standard decoding ($T_{\mathrm{draft}}=0$) Equation~3 equals $2A_A(n)+2D_V+T_{\mathrm{exp}}(r)$ exactly & modeled\\\midrule
\multicolumn{3}{@{}l}{\emph{FFN service constants (Equation~\ref{eq:app-ffn})}}\\
$\tau_\ell$ & per-layer dispatch intercept ($42.83\,\mu$s) & profile fit\\
$W_\ell$ & expert weight bytes per FFN GPU per layer & model config\\
$\mathrm{BW}, u$ & HBM bandwidth ($7.93$\,TB/s) and its efficiency ($0.89$) & spec\\
$P, u_c$ & FP8 peak ($5$\,PFLOPS) and GEMM efficiency ($0.335$) & spec / profile fit\\
$H_\ell, I_\ell, K_\ell$ & hidden width, expert intermediate width, active experts per token & model config\\
$F_{\ell,0}$ & fixed per-layer FFN overhead $\tau_\ell+W_\ell/(u\,\mathrm{BW})$, the intercept of Equation~\ref{eq:app-ffn} & modeled\\
$c_\ell$ & per-token expert cost $6H_\ell I_\ell K_\ell/(u_cP)$, the slope of Equation~\ref{eq:app-ffn} & modeled\\\midrule
\multicolumn{3}{@{}l}{\emph{Speculative decoding}}\\
$d, q$ & draft length; verified positions $q=d+1$ & config\\
$\bar a$ & mean accepted tokens per request per step & measured\\
$\kappa_\ell, D_{V,\specmode}$ & speculative attention factor and target per-microbatch auxiliary time (Table~\ref{tab:profile-inputs}) & profile\\
$T_{\mathrm{draft}}$ & graph-external draft-and-accept latency per step, $(b_A/b_p)T_{\mathrm{draft,att},V}+T_{\mathrm{draft,rest},V}$ & profile\\
$T_a^{\specmode}, T_e^{\specmode}$ & speculative attention stage $\kappa_\ell A_\ell(n)+D_{V,\specmode}/L$ / FFN stage $F_\ell(rnq)$ (Section~4.2) & modeled\\\bottomrule
\end{tabular}
\end{table*}
\begin{table}[!htbp]
\centering\small
\caption{Colocated profile inputs of the model. Standard rows use the vLLM EP4 profile; speculative rows use the profile with the same draft length. A dash marks no speculative attention factor.}\label{tab:profile-inputs}
\begin{tabular}{llrrrr}\toprule
Model & Context & $d$ & $b_p$ & $D_V$ (ms) & $\kappa$\\\midrule
MiniMax-M2.5 & 4K & 0 & 94 & 3.8310 & ---\\
MiniMax-M2.5 & 8K & 0 & 49 & 3.6400 & ---\\
MiniMax-M2.5 & 16K & 0 & 25 & 3.5070 & ---\\
Qwen3.5 & 4K & 0 & 83 & 7.5140 & ---\\
Qwen3.5 & 8K & 0 & 62 & 7.2230 & ---\\
Qwen3.5 & 16K & 0 & 45 & 7.0110 & ---\\
Qwen3 & 4K & 0 & 125 & 5.7890 & ---\\
Qwen3 & 8K & 0 & 66 & 5.5130 & ---\\
Qwen3 & 16K & 0 & 34 & 5.2220 & ---\\
MiniMax-M2.5 & 4K & 3 & 126 & 5.0788 & 1.0320\\
MiniMax-M2.5 & 4K & 7 & 109 & 6.7159 & 1.0366\\
MiniMax-M2.5 & 8K & 3 & 67 & 4.3770 & 1.0149\\
MiniMax-M2.5 & 8K & 7 & 67 & 5.1355 & 1.0273\\
MiniMax-M2.5 & 16K & 3 & 34 & 3.9178 & 0.9914\\
MiniMax-M2.5 & 16K & 7 & 34 & 4.3139 & 0.9958\\
MiniMax-M2.5 & 32K & 3 & 18 & 3.3866 & 0.9632\\
MiniMax-M2.5 & 32K & 7 & 17 & 3.7036 & 0.9668\\
Qwen3 & 4K & 3 & 151 & 9.2576 & 1.0456\\
Qwen3 & 4K & 7 & 119 & 12.1423 & 1.1510\\
Qwen3 & 8K & 3 & 88 & 6.8859 & 1.0384\\
Qwen3 & 8K & 7 & 76 & 9.3263 & 1.0899\\
Qwen3 & 16K & 3 & 38 & 5.6180 & 1.0117\\
Qwen3 & 16K & 7 & 38 & 6.7442 & 1.0618\\
Qwen3 & 32K & 3 & 23 & 5.1682 & 1.0187\\
Qwen3 & 32K & 7 & 23 & 5.8431 & 1.1504\\
\bottomrule\end{tabular}
\end{table}

\subsection{Auxiliary Latency and Calibration}
$D_V$ aggregates the auxiliary kernels of the colocated baseline: projections, normalization, routing, and the other attention-side work outside attention and the remote experts. The model fixes it at the profile value, and each of the two microbatches incurs it once per step. This approximation holds for the decode operating points studied here. Figure~5b uses this model for the \sys{} bars: attention from the wave model, $2D_V$ from the measured vLLM EP4 profile, and FFN exposure as the remainder of the predicted step, which is within $0.2\text{--}3.8\%$ of the measured step at the three 8K peaks.

The attention estimate is a wave model with effective bandwidth utilization $u=0.89$, HBM bandwidth $7.93\times10^{12}$ bytes/s, per-SM bandwidth $107\times10^9$ bytes/s, and 152 SMs. The expert-compute model uses an FP8 rate scale $P=5\times10^{15}$ operations/s, GEMM utilization $u_c=0.3347407189$, and a fixed service term $\tau=42.834448717\,\mu$s. Both $u_c$ and $\tau$ are fitted once across the twelve vLLM EP4 profiles ($\tau$: intercept of dispatch-plus-combine time against routed bytes; $u_c$: slope of expert GEMM time against FLOPs). They are model inputs, not measured hardware maxima or isolated communication costs.

The attention-wave function uses a block count $N$, bytes per block $B_{\mathrm{blk}}$, $S_{\mathrm{SM}}=152$, $w=\lfloor N/S_{\mathrm{SM}}\rfloor$, and $k=N-wS_{\mathrm{SM}}$:
\begin{equation}
 A_{\mathrm{wave}}=\frac{wS_{\mathrm{SM}}B_{\mathrm{blk}}}{u\,\mathrm{BW}}+
 \mathbf{1}_{k>0}\max\!\left(\frac{kB_{\mathrm{blk}}}{u\,\mathrm{BW}},\frac{B_{\mathrm{blk}}}{\mathrm{BW}_{\mathrm{SM}}}\right).
\end{equation}
For MiniMax and Qwen3, $N=n h_{KV}$ and $B_{\mathrm{blk}}=4d_h(c+256)$, with context $c$, head dimension $d_h=128$, and KV-head counts eight and four. The Qwen3.5 full-attention estimate uses $N=2n$ and $B_{\mathrm{blk}}=1024(c+261)$; its gated-delta estimate is $1.3677n\,\mu$s. Remote FFN service over $m$ routed tokens ($rn$ standard; $rnq$ speculative) decomposes into dispatch intercept, HBM3e expert-weight reads, and grouped-GEMM compute:
\begin{equation}
 F_\ell(m)=\tau_\ell+\frac{W_\ell}{u\,\mathrm{BW}}
                  +\frac{m\,6H_\ell I_\ell K_\ell}{u_c P}.
\label{eq:app-ffn}
\end{equation}

For hybrid Qwen3.5, the model sums 15 full-attention and 45 gated-delta layer terms and takes the maximum with FFN service within each layer type before summing; averaging the two attention types first would hide a bottleneck in one of them. The measured full-model $D_V$ is spread evenly across layers, and the residuals include the effect of that simplification. Table~\ref{tab:profile-inputs} lists the profile inputs.

\subsection{Closed-Form Throughput and Stage-Level Win Condition}
Decomposing the colocated wide-EP step into attention, auxiliary, and MoE components as $T_V = A_V(b_V) + D_V + M_V(b_V)$, and writing the two-microbatch \sys{} step identically as $\widehat{T}_A = 2A_A(b_A/2) + 2D_V + T_{\mathrm{exp}}(r)$ where $T_{\mathrm{exp}}(r) = 2\sum_\ell \max\{0,\, F_\ell(rnq) - T_{a,\ell}\}$, Equation~2 yields $S_{\allgpu} > 1$ if and only if
\begin{equation}
 \underbrace{\left(\dfrac{b_A}{b_V}\right)}_{\text{Batch gain}}
 \!\times\!
 \underbrace{\left(\dfrac{A_V(b_V)+D_V+M_V(b_V)}{2 A_A(b_A/2)+2D_V+T_{\mathrm{exp}}(r)}\right)}_{\text{Overlap gain}}
 > 1 + \frac{1}{r}.
\label{eq:app-boundary}
\end{equation}
Under a fully covered FFN stage ($T_{\mathrm{exp}}(r)=0$) and linear attention scaling, let the baseline period be $T_V=A_V+D_V+M_V$, and let $B=b_A/b_V$. Approximating $T_A=B A_V+2D_V$ gives
\begin{align}
 S&=\frac{r}{r+1}\frac{B}{B(1-d_V-m_V)+2d_V},\\
 S>1&\Longleftrightarrow m_V>\frac{1}{r+1}+\left(\frac{2}{B}-1\right)d_V.
\end{align}
This closed form holds under the stated approximation; attention-wave effects, exposed FFN service, hybrid layers, and speculative auxiliary work require the full per-layer model.

\subsection{Speculative Kernel Partition and Two-Side Decomposition}
The speculative profiles cover 16 model/context/draft combinations. The extraction identifies the main target CUDA graph and partitions its attention, expert, and auxiliary kernels ($D_{V,\specmode}$) from graph-external draft-and-accept activity ($T_{\mathrm{draft}}$). Draft attention is scaled by $b_A/b_p$ ($T_{\mathrm{draft}}\approx (b_A/b_p)T_{\mathrm{draft,att},V}+T_{\mathrm{draft,rest},V}$), where $b_p$ is the profiling run's batch, distinct from the comparison baseline's $b_V$; the remaining external activity is fixed at its profile value. Target verification already processes $q=d+1$ positions, so its measured auxiliary work $D_{V,\specmode}$ is not multiplied by $q$ again.

Separating the two sides of Equation~3 in the main paper, verification attends $q$ positions per request against the same prefix on the attention GPUs, reading each KV cache once:
\begin{equation}
 T_a^{\specmode} = \kappa_\ell A_\ell(n) + D_{V,\specmode}/L.
\label{eq:app-spec-local}
\end{equation}
On the expert side, the FFN pool serves $q\times$ more token rows ($rnq$) while reading each expert's weights once per layer. Writing $F_\ell(m)=F_{\ell,0}+m\,c_\ell$ with fixed per-layer dispatch and weight-read overhead $F_{\ell,0}=\tau_\ell+W_\ell/(u\,\mathrm{BW})$ and per-token expert GEMM cost $c_\ell=6H_\ell I_\ell K_\ell/(u_c P)$ in Equation~\ref{eq:app-ffn}, only the linear expert-GEMM term grows:
\begin{align}
 T_e^{\specmode} &= F_\ell(rnq) = T_e + \Delta F_\ell,\nonumber\\
 \Delta F_\ell &= (q-1)\,rn\,c_\ell = (q-1)\frac{rn\,6H_\ell I_\ell K_\ell}{u_c P}.
\label{eq:app-headroom}
\end{align}
The split is reproducible at the kernel level but does not label every draft, sampling, acceptance, or synchronization kernel individually, and summed CUDA time need not equal the critical-path time of concurrent streams; the speculative errors in Appendix~\ref{app:model} measure how well this compact split works.

\section{Complete Speculative Comparisons}
\label{app:spec}
Table~\ref{tab:spec-best} lists the best measured configurations for standard decoding and for draft lengths three and seven on \mm{} and \qwen{} at 4K--32K. Draft and target models both run, and acceptance is held near 1.8 useful tokens per request per step; run-level calculations use the measured averages. Batch and topology may change across modes, and each throughput is normalized by the GPUs that configuration allocates.

All-GPU and attention-only speedups differ by the FFN-node factor $r/(r+1)$ of the AFD configuration; the attention-only value describes the attention pool and does not treat FFN GPUs as free. Self-speedup compares each engine with its own standard-decode reference; speedups over wide EP compare the two engines in the same mode. \mm{} 32K at $d=3$ has no valid EP64 run, so its wide-EP peak is not bracketed (EP32 beats EP16); an asterisk marks this cell.
\begin{table*}[!htbp]
\centering\small
\caption{Best measured configurations for standard ($d=0$) and speculative ($d=3,7$) decoding (tokens/s/GPU, all GPUs counted). Acceptance is held near 1.8 useful tokens per step. $^{*}$MiniMax-M2.5 32K, $d=3$: EP64 has no valid run, so the wide-EP peak is not bracketed.}\label{tab:spec-best}
\begin{tabular}{lll lrlrrr}\toprule
Model & Context & $d$ & AFD & tok/s/GPU & EP & tok/s/GPU & $S_{all}$ & $S_{att}$\\\midrule
MiniMax-M2.5 & 4K & 0 & 9:1 & 4118.1 & EP32 & 4098.0 & 1.005 & 1.117\\
MiniMax-M2.5 & 4K & 3 & 4:1 & 5213.2 & EP16 & 4849.4 & 1.075 & 1.344\\
MiniMax-M2.5 & 4K & 7 & 3:1 & 3770.2 & EP8 & 3551.7 & 1.062 & 1.415\\
MiniMax-M2.5 & 8K & 0 & 15:1 & 2372.1 & EP32 & 2317.0 & 1.024 & 1.092\\
MiniMax-M2.5 & 8K & 3 & 5:1 & 3302.6 & EP16 & 3238.4 & 1.020 & 1.224\\
MiniMax-M2.5 & 8K & 7 & 4:1 & 2544.2 & EP16 & 2497.4 & 1.019 & 1.273\\
MiniMax-M2.5 & 16K & 0 & 17:1 & 1234.2 & EP32 & 1208.0 & 1.022 & 1.082\\
MiniMax-M2.5 & 16K & 3 & 8:1 & 1903.9 & EP16 & 1758.2 & 1.083 & 1.218\\
MiniMax-M2.5 & 16K & 7 & 6:1 & 1517.5 & EP16 & 1520.9 & 0.998 & 1.164\\
MiniMax-M2.5 & 32K & 0 & 17:1 & 603.6 & EP32 & 600.7 & 1.005 & 1.064\\
MiniMax-M2.5 & 32K & 3 & 14:1 & 1021.8 & EP32$^{*}$ & 999.2 & 1.023$^{*}$ & 1.096$^{*}$\\
MiniMax-M2.5 & 32K & 7 & 8:1 & 855.0 & EP16 & 862.7 & 0.991 & 1.115\\
Qwen3 & 4K & 0 & 5:1 & 4256.4 & EP16 & 4004.0 & 1.063 & 1.276\\
Qwen3 & 4K & 3 & 2:1 & 4439.1 & EP8 & 3912.5 & 1.135 & 1.702\\
Qwen3 & 4K & 7 & 2:1 & 2699.4 & EP4 & 2443.7 & 1.105 & 1.657\\
Qwen3 & 8K & 0 & 7:1 & 2624.3 & EP16 & 2389.0 & 1.098 & 1.255\\
Qwen3 & 8K & 3 & 3:1 & 3064.2 & EP16 & 2794.4 & 1.097 & 1.462\\
Qwen3 & 8K & 7 & 2:1 & 2122.8 & EP8 & 1899.5 & 1.118 & 1.676\\
Qwen3 & 16K & 0 & 11:1 & 1451.6 & EP32 & 1240.0 & 1.171 & 1.277\\
Qwen3 & 16K & 3 & 5:1 & 1892.8 & EP8 & 1696.8 & 1.116 & 1.339\\
Qwen3 & 16K & 7 & 3:1 & 1393.2 & EP8 & 1306.6 & 1.066 & 1.422\\
Qwen3 & 32K & 0 & 16:1 & 735.8 & EP32 & 713.0 & 1.032 & 1.096\\
Qwen3 & 32K & 3 & 8:1 & 1072.0 & EP16 & 1006.3 & 1.065 & 1.198\\
Qwen3 & 32K & 7 & 6:1 & 806.5 & EP16 & 740.9 & 1.089 & 1.270\\
\bottomrule\end{tabular}
\end{table*}

\section{Ablations and CAD}
\label{app:ablation}

\begin{figure*}[t]
\centering
\includegraphics[width=\textwidth]{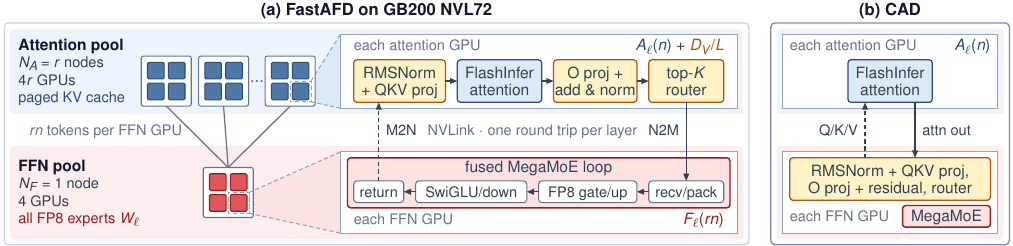}
\vspace{-0.65em}
\caption{\textbf{Operator split boundary comparison.} \textbf{(a)} \sys{} keeps projections and routing on the attention pool around an off-path \texttt{N2M}/\texttt{M2N} round trip to fused \texttt{MegaMoE}. \textbf{(b)} Core Attention Disaggregation (CAD) moves projections to the FFN pool, placing its round trip on the attention critical path and shrinking the attention stage.}
\vspace{-0.55em}
\label{fig:app-cad-arch}
\end{figure*}

\paragraph{Runtime ablations.} The \mm{} 8K, 7:1, batch-76 ablation uses preset \texttt{afd-cell-v3.4.3} (Table~\ref{tab:presets}). Base, single-microbatch, and no-cross-step-overlap variants have wall-clock periods 29.893, 39.880, and 43.820\,ms, respectively, and throughput of 2224.6, 1667.5, and 1517.6 tokens/s/GPU. The estimator is the inverse slope of decode-step number against the log timestamps. Internal cell means are 29.906, 39.874, and 40.483\,ms; the last is not interchangeable with the wall-clock number because the disabled-overlap loop changes the timing scope.

Table~\ref{tab:ablation} lists every ablation leg behind Figure~7a and c, and Table~\ref{tab:microbatch} the microbatch sweep behind Figure~7b. On Qwen3, the non-fused \texttt{DeepEP}+\texttt{DeepGEMM} variant needs more than 12\,GiB of extra memory per attention GPU: at memory ratio 0.88 ($b_A=102$), CUDA-graph capture of its dispatch runs out of memory on all 28 attention GPUs, so both legs of the pair use ratio 0.80 ($b_A=92$); the \texttt{MegaMoE} run at the Table~\ref{tab:best} configuration ($b_A=106$, ratio 0.91) reproduces the table within $0.06\%$. Non-fused wall-clock periods exceed the internal step means by about 6\% (53.21 vs.\ 50.02\,ms on \mm{}, 67.70 vs.\ 63.79\,ms on Qwen3) because its steps serialize host work; the figure uses wall-clock periods for every leg.
\begin{table*}[!htbp]
\caption{Runtime ablation legs behind Figure~7a and c: measured wall-clock periods, one run per row. MR is the run's KV memory ratio.}
\label{tab:ablation}\centering\small
\setlength{\tabcolsep}{3pt}
\begin{tabular}{@{}llrrr@{}}\toprule
Config & Variant & $b_A$ & $T$ (ms) & tok/s/GPU\\\midrule
MiniMax-M2.5 8K 7:1 & MegaMoE (full \sys{}) & 76 & 29.813 & 2230.6\\
 & DeepEP+DeepGEMM & 76 & 53.209 & 1249.8\\
 & full \sys{} (second session) & 76 & 29.893 & 2224.6\\
 & one microbatch & 76 & 39.880 & 1667.5\\
 & no cross-step overlap & 76 & 43.820 & 1517.6\\
\midrule
Qwen3 8K 7:1 & MegaMoE, reported batch (MR 0.91) & 106 & 35.322 & 2625.8\\
 & DeepEP+DeepGEMM, MR 0.88 & 102 & \multicolumn{2}{c}{out of memory}\\
 & MegaMoE, MR 0.80 & 92 & 33.542 & 2400.0\\
 & DeepEP+DeepGEMM, MR 0.80 & 92 & 67.701 & 1189.1\\\bottomrule
\end{tabular}
\end{table*}

\begin{table}[!htbp]
\caption{Microbatch-count sweep behind Figure~7b on \mm{} 8K: measured wall-clock periods (ms) at $b_A=76$ ($b_A=75$ with three microbatches, so every microbatch keeps a captured decode graph). The repeat is an independent 3:1 session on other nodes; the last column uses $b_A=72$. The bottom row is the reported configuration with two microbatches.}
\label{tab:microbatch}\centering\small
\setlength{\tabcolsep}{4pt}
\begin{tabular}{@{}lrrrr@{}}\toprule
Microbatches & 3:1 & 3:1 repeat & 7:1 & 7:1, $b_A=72$\\\midrule
1 & 37.770 & 37.720 & 39.902 & 38.214\\
2 & 29.607 & 29.605 & 29.873 & 28.825\\
3 & 35.061 & 35.033 & 35.139 & 34.774\\
4 & 36.699 & 36.634 & 38.797 & 38.649\\
\midrule
Reported cell & 29.688 & 29.688 & 29.827 & ---\\\bottomrule
\end{tabular}
\end{table}

\paragraph{CAD sweep.} CAD moves the QKV/output projections and expert execution to the FFN node while keeping the attention core, Q/K normalization, and KV writes on the attention nodes. A token crosses this boundary before and after attention, so Equation~3 does not model CAD latency. Table~\ref{tab:cad} reports the complete 15-cell CAD sweep on \mm{} across 4K--16K. CAD runs use a small-tile \texttt{MegaMoE} setting that favors CAD relative to the AFD tables, so the comparison is conservative for \sys{}. The CAD optimum is bracketed on both sides at 2:1--4:1 and reaches $0.675\text{--}0.782\times$ of \sys{}'s best per-GPU throughput. All 15 CAD cells produce output token-identical to their \sys{} reference runs (the first 1, 16, and 64 tokens and the full output of every sampled prompt); speculative modes were not compared.

\begin{table}[!htbp]
\caption{Complete 15-cell CAD measurements on \mm{} across 4K--16K. Each context brackets its throughput peak on both sides (\texttt{kvfull} maximizes per-GPU KV batch; \texttt{aligned} matches the selected AFD batch $b_A$). Throughput counts all GPUs.}
\label{tab:cad}\centering\small
\setlength{\tabcolsep}{4pt}
\begin{tabular}{llrrrr}\toprule
Ctx & Scope & A:F & $b$ & $T$ (ms) & tok/s/GPU\\\midrule
4K & kvfull  & 1:1 & 154 & 31.315 & 2458.9\\
4K & kvfull  & \textbf{2:1} & 154 & 36.918 & \textbf{2781.0}\\
4K & kvfull  & 3:1 & 154 & 42.285 & 2731.4\\
4K & kvfull  & 4:1 & 154 & 45.889 & 2684.7\\\midrule
8K & kvfull  & 2:1 &  82 & 32.796 & 1666.9\\
8K & kvfull  & \textbf{3:1} &  82 & 35.839 & \textbf{1716.0}\\
8K & aligned & 3:1 &  76 & 35.618 & 1600.3\\
8K & kvfull  & 4:1 &  82 & 39.677 & 1653.3\\
8K & kvfull  & 5:1 &  82 & 42.157 & 1620.9\\\midrule
16K & kvfull  & 3:1 &  42 & 32.668 &  964.2\\
16K & kvfull  & \textbf{4:1} &  42 & 34.827 &  \textbf{964.8}\\
16K & aligned & 4:1 &  38 & 34.436 &  882.8\\
16K & kvfull  & 5:1 &  42 & 37.187 &  941.2\\
16K & kvfull  & 6:1 &  42 & 39.180 &  918.8\\
16K & kvfull  & 7:1 &  42 & 41.317 &  889.5\\\bottomrule
\end{tabular}
\end{table}

\clearpage
\section{Reproducibility}
\label{app:repro}
Table~\ref{tab:presets} lists the run presets behind every measured standard-decode cell. Each run records its preset, rendered environment, and code version: all \mm{} and \qwen{} table runs use one \sys{} version and all \hybrid{} runs another. vLLM runs use v0.31.0 with the DeepEP low-latency all-to-all, \texttt{-{}-max-num-batched-tokens 1024}, and \texttt{-{}-block-size 64}; on \hybrid{}, a startup hook enables the DeepGEMM block-FP8 linears for which stock v0.31.0 falls back to CUTLASS on Blackwell. \revisionNov{The \hybrid{} vLLM rows use v0.33.0 with the same settings plus \texttt{-{}-no-enable-prefix-caching} and the Triton linear-attention prefill: with prefix caching on, vLLM keeps extra linear-attention state blocks for every request, which \sys{} never allocates. Without prefix caching the client cannot warm the prompts in advance, so the scheduler prefills every request before the first decode step, and all requests then decode together inside the measured window. We use this configuration because it gives vLLM its best measured performance on \hybrid{}, so the comparison and analysis are against the strongest vLLM baseline.} Both engines decode attention with the same TRT-LLM-gen FMHA kernels. Both also run the MoE experts as FP8 grouped GEMMs with E8M0 scales: vLLM through its DeepGEMM FP8 MoE backend, and \sys{} by rounding each checkpoint's $128\times128$ block scales up to the per-32 UE8M0 format of the Mega MoE kernel once at load.

\begin{table}[!htbp]
\caption{Run presets behind the measured standard-decode cells (62 \sys{}, 60 vLLM). OSL: output tokens per request.}
\label{tab:presets}\centering\scriptsize
\setlength{\tabcolsep}{3pt}
\begin{tabular}{@{}lrl@{}}\toprule
Preset & Cells & OSL\\\midrule
\multicolumn{3}{@{}l}{\emph{\sys{}, MiniMax-M2.5 and Qwen3}}\\
\texttt{afd-cell-v3.4} & 10 & 1024\\
\texttt{afd-cell-v3.4.1} & 25 & 1024\\
\texttt{afd-cell-v3.4.2} & 3 & 1024\\
\texttt{afd-cell-v3.4.3} & 9 & 1024\\
\midrule \multicolumn{3}{@{}l}{\emph{\sys{}, Qwen3.5}}\\
\texttt{afd-cell-q5balign-v2} & 13 & 1024\\
\texttt{afd-cell-q5balign-v2.1} & 2 & 1024\\
\midrule \multicolumn{3}{@{}l}{\emph{vLLM v0.31.0, all models\revisionNov{ (\hybrid{}: v0.33.0)}}}\\
\texttt{vllm-ep-v1.6} & 9 & 1024\\
\texttt{vllm-ep-v1.7.1} & 35 & 1024\\
\texttt{vllm-ep-v1.7.2} & 4 & 1024\\
\texttt{vllm-ep-v1.7.4} & 4 & 1024\\
\texttt{vllm-ep-v1.7.5} & 4 & 1024\\
\texttt{vllm-ep-v1.7.8} & 2 & 1024\\
\texttt{vllm-ep-v1.7.11} & 2 & 1024\\\bottomrule
\end{tabular}
\end{table}

\end{document}